\documentclass[aps,pra,onecolumn,floatfix,superscriptaddress,showkeys,10pt]{revtex4-2}
\usepackage{mathbbol}
\usepackage{amsmath,amssymb}
\usepackage{tikz}
\usepackage{centernot}
\usepackage[utf8]{inputenc}
\usepackage[english]{babel}
\usepackage{amsfonts}
\usepackage{amsthm}
\usepackage{mathrsfs}
\usepackage{mathtools}
\usepackage[nice]{nicefrac}
\usepackage{amsmath}
\usepackage{xfrac}
\usepackage{tabularx}
\usepackage{physics,subfigure}
\usepackage{tikz}
\usetikzlibrary{positioning}
\usepackage{tkz-berge}
\usetikzlibrary{arrows}
\usetikzlibrary{graphs,graphs.standard,fit,shapes.geometric,calc,hobby}
\usepackage[colorlinks=true,hyperfootnotes=true,breaklinks=true,citecolor=red,urlcolor=blue,linkcolor=blue]{hyperref}

\usepackage{relsize}

\begin{document}
%
\title{Quantum metrology of electric and magnetic dipole moments: ultimate limits and optimal regimes}
%
\author{Simone Cavazzoni}
\email{simone.cavazzoni@unimore.it}
\affiliation{Dipartimento di Scienze Fisiche, Informatiche e Matematiche,  Universit\`{a} di Modena e Reggio Emilia, I-41125 Modena, Italy}
\author{Paolo Bordone}
\email{paolo.bordone@unimore.it}
\affiliation{Dipartimento di Scienze Fisiche, Informatiche e Matematiche, Universit\`{a} di Modena e Reggio Emilia, I-41125 Modena, Italy}
\affiliation{Centro S3, CNR-Istituto di Nanoscienze, I-41125 Modena, Italy}
\author{Matteo G. A. Paris}
\email{matteo.paris@fisica.unimi.it}
\affiliation{Quantum Technology Lab, Dipartimento di Fisica {\em Aldo Pontremoli}, Universit\`{a} degli Studi di Milano, I-20133 Milano, Italy}
\affiliation{INFN, Sezione di Milano, I-20133 Milano, Italy}
\date{\today}
\begin{abstract}
The characterization of electric and magnetic dipole moments (EDM and MDM) in quantum systems is central to fundamental physics and quantum sensing. While EDM searches provide powerful probes of CP violation within and beyond the Standard Model, precise MDM estimation is crucial for high-precision magnetometry and the development of quantum sensors. In this work, we address the ultimate precision limits for separate and simultaneous estimation of both dipole moments in a generic two-level system coupled to electromagnetic fields. We analyze three classes of quantum probes/strategies: unitary and depolarizing dynamics, and thermal equilibrium states. For each, we derive the quantum Fisher information (matrix), identify optimal probes, and determine the ideal operating conditions, such as evolution times and temperatures, that maximize estimation precision. We further assess the compatibility and sloppiness of the statistical models, showing that orthogonal dipole moments configurations enable joint estimation of EDM and MDM, whereas parallel configurations are intrinsically sloppy, permitting only the estimation of a single parameter combination. Our results provide a unified metrological framework for estimation schemes ranging from neutron EDM searches to molecular magnetometry, and highlight the distinct roles of coherence, noise, and thermalization in multiparameter quantum sensing of dipole moments.

\end{abstract}
\maketitle

\section{Introduction} 
\label{sec:I}

The precise characterization of the electromagnetic properties of a generic quantum system is based 
on its interaction with external fields, fully captured by its electric and magnetic dipole moments 
(EMDM). If a system has a magnetic moment, its Hamiltonian contains a Zeeman contribution. If it also 
has an electric dipole moment, then there is the additional Stark-Lo Surdo term. The resulting 
parameter-dependent Hamiltonian thus encodes the complete coupling of the system’s dipole moments 
to the electromagnetic field, forming the theoretical foundation for their study.

In high-energy physics, the experimental search for permanent electric dipole moments has become a powerful probe of fundamental symmetries, as a non-zero EDM would signal the violation of both time-reversal ($T$) and the combined charge-parity ($CP$) symmetry \cite{baker2006improved,pospelov2014ckm,abel2020measurement}. The neutron, despite being electrically neutral, is a central subject of these investigations because its internal quark structure can, in principle, host both magnetic and electric dipole moments, intimately linked to $CP$ violation \cite{altarev2012next,piegsa2013new,pendlebury2015revised,martin2020current}. A measurable neutron EDM would not only provide unambiguous evidence for new sources of $CP$ violation beyond the Standard Model but is also considered essential for explaining the observed matter-antimatter asymmetry of the universe, potentially through baryogenesis mechanisms. Driven by the same motivations, the search has been extended to the electron's EDM, grounded in the theorem that any non-zero permanent EDM of an elementary particle requires the violation of both parity and time-reversal symmetries \cite{hudson2002measurement}. These experiments either target the bare electron or are based 
on the enhanced sensitivity offered by specific heavy polar molecules \cite{denis2019enhancement,augenbraun2020laser,ng2022spectroscopy,caldwell2023systematic}. The molecular approach directly exploits the concept of Schiff moments, which effectively describe the screened EDM of a nucleus or atom within a complex system \cite{engel2025nuclear,hubert2022electric,flambaum2020electric}. Key systems for this class of precision measurements include nuclei of $^{199}\text{Hg}$ and $^{225}\text{Ra}$, as well as molecules such as thorium monoxide ($ThO$) and ytterbium fluoride ($YbF$) \cite{graner2017erratum,parker2015first,bishof2016improved,acme2018improved,hudson2002measurement,lim2018laser}.

The interest in EMDM goes far beyond the interest in high energy physics \cite{gubskaya2002total,minkin2012dipole,kotochigova2003ab} and extend to quantum sensing and metrology \cite{albarelli2023fundamental,suzuki2015parameter,suzuki2016explicit}. Indeed, together with the electric dipole moment, its magnetic counterpart (MDM) is crucial in  sensing of magnetic fields (magnetometry) \cite{troiani2018universal,gusarov2023optimized}. Specifically, once the dipole moments are precisely estimated, the system itself becomes an effective and precise quantum sensor. The simplest—and perhaps most paradigmatic—example of this approach is the use of a single spin systems for electromagnetic field sensing \cite{boixo2007generalized,de2011single,pang2014quantum,rodriguez2018probing}. Indeed, the metrological applications of such model in sensing devices has been extensively investigated \cite{forghieri2023quantum,fanucchi2025giant,secchi2025hole}, especially in quantum magnetometry, as it promises better performances than the classical counterparts \cite{brask2015improved}. Even if the estimation of solely the magnetic field without the electric component has been studied before in single spin systems and beyond \cite{wildermuth2005microscopic,baumgratz2016quantum,albarelli2017ultimate,pang2017optimal,liu2020quantum,rossi2020noisy,yang2022variational,montenegro2022sequential}, the aim of the present work is to generalize the framework to a multiparameter estimation scenario, investigating also the fundamental implications of the presence of electric and magnetic dipole moments in subatomic particles. The present work thus generalizes the standard magnetometry model by fully accounting for the simultaneous presence of both electric and magnetic dipole moments, and analyze the fundamental implications of the model. By examining the coupled EDM-MDM system under both ideal and realistic, noisy conditions, we provide a comprehensive framework for the characterization of the full dipole-moment tensor of any quantum system.

This manuscript is structured as follow. In Sec.\ref{sec:ToE} we present the basic theory of EMDM, from the Hamiltonian to the definition of the system in pure and non pure state formalism. The description mainly focus on Hamiltonian and Lindblad time evolution to completely characterize the quantum dynamics of the system and on the thermal state description as a paradigmatic example of non pure static state description of quantum systems. In Sec.\ref{sec:LQET} we provide the basis of local estimation theory, with the detailed formalism of single and multi parameter metrology. Specifically we will focus on the (quantum) Fisher information formalism setting the ultimate bound of precision achievable in estimation procedures of dipole moments. In Sec.\ref{sec:RaD} we illustrate our results providing formulas for the optimal state to adopt in experimental characterization of both electric and magnetic dipole moment in all the conditions we have considered. In Sec.\ref{sec:FuIm} we discuss the fundamental implications of our results. Finally, in Sec.\ref{sec:C} we draw our conclusions. Additional mathematical material about suitable reference frames for dipole estimation or the interaction of quantum systems with electromagnetic field can be found in Appendix \ref{sec:AM1} and \ref{sec:AM2}.

\section{Dipole Moments}
\label{sec:ToE}
In this Section, we provide the basic tools for the description of dipole moments of elementary particles and molecular compounds. Independently of the nature of the system, a quantum system with dipole moments, interacts with the electromagnetic via the following Hamiltonian
\begin{equation}
    \label{eq:EDM_Hamiltonian}
    \mathcal{H}= \mu\, \Vec{B}\cdot\Vec{\sigma} + d\,  \Vec{E}\cdot\Vec{\sigma}
\end{equation}
defined in a Hilbert space $\mathscr{H}=\{\ket{0},\ket{1}\}$, and where $\vec{\sigma}$ is the vector 
of Pauli matrices, which reads:
\begin{equation}
    \label{eq:Pauli_Vector}
    \vec{\sigma}=\left[ \left(\begin{array}{cc}
       0  &  1\\
         1& 0
    \end{array}\right), \left(\begin{array}{cc}
       0  &  -i\\
         i& 0
    \end{array}\right),\left(\begin{array}{cc}
       1  &  0\\
         0& -1
    \end{array}\right) \right]^T.
\end{equation}
The term with the magnetic dipole $\mu$ describes the Zeeman Effect and the term with the electric dipole $d$ indicates the Stark-Lo Surdo contribution, and together they describe the coupling of the system with the magnetic and electric field $\Vec{B}$ and $\Vec{E}$. Without loss of generality we can rewrite the Eq.\eqref{eq:EDM_Hamiltonian} as 
\begin{equation}
    \label{eq:EDM_Hamiltonian_red}
    \mathcal{H}_{r}= \mu\, B_x \, \sigma_x \pm \mu\, B_z \, \sigma_z + d\, E_z \,\sigma_z ,
\end{equation}
since a suitable coordinate frame can always be defined where the electric and magnetic dipole moments and the fields have components in only two directions (see Appendix \ref{sec:AM1}). Further simplifications can be made upon considering the specific geometry of the problem. In nucleons, nuclei or elementary particles alone, such as the neutron or the electron, the most general Hamiltonian can be written as
\begin{equation}
    \label{eq:EDM_Hamiltonian_red_parallel}
    \mathcal{H}_{r,\parallel}= \pm \mu\, B_z \, \sigma_z + d\, E_z \,\sigma_z \, .
\end{equation}
In fact, considering a particle with spin $s=\nicefrac{1}{2}$ both moments must lie along the spin direction \footnote{This condition corresponds to considering the component $\mu\, B_x \, \sigma_x=0$ in Eq.\eqref{eq:EDM_Hamiltonian_red}}, since the spin is the only vector that can orient the system \cite{commins2007electric}. Nonetheless, the components of the electric and magnetic dipole moments in every quantum system may vary, as they intrinsically depends on the symmetry of the system. Indeed, another important configuration occurs when the magnetic and electric fields are orthogonal, and the Hamiltonian of Eq.\eqref{eq:EDM_Hamiltonian_red} is reduced to
\begin{equation}
    \label{eq:EDM_Hamiltonian_red_orthogonal}
    \mathcal{H}_{r,\perp}= \mu\, B_x \, \sigma_x + d\, E_z \,\sigma_z \, .
\end{equation}
The  model may be applied to several physical systems, as those aimed to sensing CP-violating permanent EDM of the electron in metastable states of heavy polar molecules such as $ThO$ \cite{vutha2011magnetic} or in $YbF$ compounds \cite{hudson2002measurement}.

\subsection{Dynamics of dipole moments}
The time evolution of a generic initial state $\ket{\psi (0)} = \cos{\left(\frac{\alpha}2\right)} \ket{0} + e^{i\kappa}\sin{\left(\frac{\alpha}2\right)} \ket{1} \nonumber$ with $\alpha \in [0,\pi]$ and $\kappa\in\left[0,2\pi\right)$ is governed by the Hamiltonian in Eq.\eqref{eq:EDM_Hamiltonian_red}. The evolution 
operator may be written as (see Appendix \ref{sec:AM2})
\begin{align}
    \label{eq:time_evolution_operator}
    &\mathcal{U}_{r}(t) = e^{-i\mathcal{H}_{r}t}  \\ 
    &= \left[ \cos{(\omega t)} -\frac{i \zeta \sin{(\omega t)}}{\omega} \right] \ket{0} \bra{0} + \left[ \frac{-i \chi \sin{(\omega t)}}{\omega} \right] \ket{0} \bra{1} \nonumber \\ 
    &+ \left[ \frac{-i \chi \sin{(\omega t)}}{\omega} \right] \ket{1} \bra{0} + \left[ \cos{(\omega t)} +\frac{i \zeta  \sin{(\omega t)}}{\omega} \right] \ket{1} \bra{1} \nonumber
\end{align}
where $\chi=\mu B_x$, $\zeta=(\pm \mu \, B_z + d \, E_z)$ and $\omega = \sqrt{\chi^2 + \zeta^2}$. The time evolution 
may be alternatively obtained by decomposing the initial state on the eigenstates of the Hamiltonian n Eq.\eqref{eq:EDM_Hamiltonian_red}, which are given by
\begin{align}
    \label{eq:eigenstate_n3cell}
     \mathcal{E}_{-}= -\omega, \quad \ket{\phi_{-}} & = \frac{\left( -\chi \ket{0} + \left( \omega + \zeta \right) \ket{1} \right)}{\sqrt{2\omega(\omega+\zeta)}}    = \phi_{0-} \ket{0} + \phi_{1-} \ket{1} \nonumber \\
     \mathcal{E}_{+}= +\omega, \quad \ket{\phi_{+}} & = \frac{\left( (\omega + \zeta) \ket{0} + \chi \ket{1} \right)}{\sqrt{2\omega(\omega+\zeta)}} 
     = \phi_{0+} \ket{0} + \phi_{1+} \ket{1}
\end{align}
By writing
\begin{align}
    \label{eq:initial_state_dec}
    \ket{\psi(0)} = \bra{\phi_+}\ket{\psi(0)} \ket{\phi_+} + \bra{\phi_-}\ket{\psi(0)} \ket{\phi_-}
\end{align}
the evolved state is given by
\begin{align}
    \label{eq:initial_state_time_evolution}
    \ket{\psi(t)} &= e^{-i\omega t}\bra{\phi_+}\ket{\psi(0)} \ket{\phi_+} \nonumber \\
    &+ e^{i\omega t}\bra{\phi_-}\ket{\psi(0)} \ket{\phi_-} \,.
\end{align}
The time evolution encodes information about both dipole moments, because of the dependence of $\omega$ and of the eigenvectors $\ket{\phi_{\pm}}$ on $\mu$ and $d$. Explicit calculations for both the parallel or orthogonal dipole configurations will be presented in Sec.\ref{sec:RaD}. In realistic physical implementations the presence of interactions with the environment and fluctuations leads to a noisy (non-unitary) evolution of a density matrix \cite{daffer2004depolarizing} in the form $\rho = \left( a \ket{0} \bra{0} + b \ket{0} \bra{1} + b^{*} \ket{1} \bra{0} + d \ket{1} \bra{1} \right) $. The same matrix can be written using the Bloch formalism in the form
\begin{equation}
    \label{eq: bloch rho}
    \rho = \frac{1}{2} \left( \mathbb{I} + \vec b  \cdot  \vec \sigma \right)
\end{equation}
with $\vec b$ which is the Bloch vector of the matrix and it is defined as
\begin{equation}
\label{eq:r}
    \vec b = (b_{x},b{y},b_{z})^{T} = (\Tr[\rho \, \sigma_{x}], \Tr[\rho \, \sigma_{y}] , \Tr[\rho \, \sigma_{z}])^{T} \,.
\end{equation}

For the sake of generality, we here consider isotropic noise, with the dynamics described by the Lindblad master equation \cite{manzano2020short}
\begin{equation}
    \label{eq:depolarized_evolution}
    \partial_{t} {\rho}(t) = -i\left[ \mathcal{H},{\rho}(t) \right] - \frac{\gamma}{4} \sum_{i=1}^{3} \left( \sigma_{i} \rho(t) \sigma_{i}^{+} - \rho(t) \right),
\end{equation}
which consists in the sum of the Hamiltonian evolution and a depolarizing term, which progressively drives the initial state towards the maximally mixed state. Under this assumption the information about the dipole moments carried by the state of the system $\rho(t)$ is progressively degraded up to a situation at $t \rightarrow \infty$ in which the state of the system is maximally mixed and do not contains any more information about the Hamiltonian evolution. As we will see in the following sections, this condition lead to the appearance of an optimal time evolution, maximizing the precision of the estimation of the dipole moments $\mu$ and $d$.

\subsection{Equilibrium states}
In stationary condition, a quantum system in equilibrium with its environment is described by thermal Gibbs state \cite{kammerlander2016coherence,marvian2022operational,razavian2019quantum}. In such conditions, the density matrix 
of a two level system is diagonal in the Hamiltonian eigenbasis and in the form
\begin{align}
    \label{eq:thermal_rho}
    \rho_{th} = \frac{1}{\mathcal{Z}} \left( e^{-\beta\mathcal{E}_{-}} \ket{\phi_{-}}\bra{\phi_{-}} + e^{-\beta\mathcal{E}_{+}} \ket{\phi_{+}}\bra{\phi_{+}} \right) 
\end{align}
where $\beta=1/T$ is the inverse temperature (we use natural units and the Boltzmann constant $k_{B}$ is set to unit), as and $\mathcal{Z}=\Tr[e^{-\beta \mathcal{H}}]$ is the partition function. This means that even in stationary condition and without a  control over the environment, the system still encodes information about the dipole parameters and the components of the electromagnetic field, since its statistical description depends both on $\omega$ and $\ket{\phi_{\pm}}$.


It is thus possible to encode and then extract information about the dipole moments of the system in both 
cases, by dynamical encoding  or in stationary conditions. This is obtained by measuring an observable. The 
choice of the optimal observable, allowing one to achieve the maximum precision, may be performed using 
tools from local quantum estimation theory, which will be briefly reviewed in the next Section.

\section{Local Quantum Estimation Theory}
\label{sec:LQET}

In the field of local quantum estimation the main quantities to analyze are the classical Fisher information (FI) and its quantum counterpart (QFI) \cite{paris2009quantum}. Defined as the supremum of the FI, the QFI represent the maximal amount of information that a measurement can extract from a system in analysis. On the other hand, the FI is related to the information extracted through a specific measurement procedure. Both these quantities can be related to one or more parameters $\boldsymbol{\lambda}$ that define the system (in the present paper $\boldsymbol{\lambda}=\{ \mu, d \}$). For the characterization of a single parameter $\lambda$ the Fisher Information of a variable $\lambda$ obtained through a measurement related to a discrete variable $\Pi$ is given by
\begin{equation}
\label{eq:fisher_information}
\mathcal{F}_{\Pi}(\lambda) = \sum_{\Pi} \dfrac{\left| \partial_{\lambda} \rho \left( \Pi \vert \lambda \right) \right|^{2}}{\rho\left( \Pi \vert \lambda \right)},
\end{equation}
where $ \rho \left( \Pi \vert \lambda \right) $ is the conditional probability density. The QFI is defined through the symmetric logarithmic derivative $\mathcal{L}$, as
\begin{equation}
    \label{eq:general_qfi}
    \mathcal{I}(\lambda) = Tr[\rho_{\lambda}\mathcal{L}^{2}_{\lambda}],
\end{equation}
with $\rho$ density matrix of the system and with $\mathcal{L}$ defined as 
\begin{equation}
    \label{eq:sld}
    \partial_{\lambda} \rho_{\lambda} = \frac{ \mathcal{L}_{\lambda}\rho_{\lambda} + \rho_{\lambda} \mathcal{L}_{\lambda} }{2}.
\end{equation}
For any pure state, the symmetric logarithmic derivative (SLD) can be obtained as
\begin{equation}
    \label{eq:SLD pure state}
    \mathcal{L}_{\lambda} = 2 \left( \vert \psi_{\lambda} \rangle \langle \partial_{\lambda} \psi_{\lambda} \vert  + \vert \partial_{\lambda} \psi_{\lambda} \rangle\langle  \psi_{\lambda} \vert \right)
\end{equation}
and this leads to a QFI in the form
\begin{equation}
    \label{eq:qfi_pure_state}
    \mathcal{I}(\lambda) = 4 \left[ \abs{\partial_{\lambda} \psi_{\lambda}}^{2} - \abs{\bra{\psi_{\lambda}}\ket{\partial_{\lambda} \psi_{\lambda}}}^2 \right]\ .
\end{equation}
As a consequence of Eq.\eqref{eq:sld} the spectral decomposition of the SLD operator provides the optimal basis for a measurement procedure of the unknown parameters. The FI and QFI intrinsically define the variance of the measurement procedure through the Cramer-Rao bound as
\begin{equation}
    \label{eq:qvar}
	\hbox{Var}(\lambda) \geq \frac{1}{M\mathcal{F}_{\Pi}(\lambda)} \geq \frac{1}{M\mathcal{I}(\lambda)},
\end{equation}	
where $M$ is the number of measurements.
Extending the concept to a multiparameter joint estimation of system's parameters, the FI becomes a matrix, and reads
\begin{equation}
\label{eq:fisher_information_matrix}
\mathcal{F}_{\Pi}(\boldsymbol{\lambda})_{n,m} = \sum_{\Pi} \dfrac{ \partial_{n} \rho \left( \Pi \vert \boldsymbol{\lambda} \right) \partial_{m} \rho \left( \Pi \vert \boldsymbol{\lambda} \right)} {\rho\left( \Pi \vert \boldsymbol{\lambda} \right)},
\end{equation}
where the derivative $\partial_{n}$ and $\partial_{m}$ refers to the derivative with respect to the parameter $\lambda_{n}$ and $\lambda_{m}$. As its classical counterpart, also the QFI becomes a matrix, defined as
\begin{equation}
    \label{eq:quantum_fisher_information_matrix}
    \mathcal{I}(\boldsymbol{\lambda})_{n,m} = Tr\left[\rho_{\boldsymbol{\lambda}}\frac{\mathcal{L}_{n}\mathcal{L}_{m}+\mathcal{L}_{m}\mathcal{L}_{n}}{2}\right].
\end{equation}
which for pure states can be obtained generalizing Eq.\eqref{eq:qfi_pure_state} as
\begin{equation}
    \label{eq: QFI multi Pure}
    {\mathcal{I}}_{nm}\boldsymbol{(\lambda)} =  4 \Re{\left[ \bra{\partial_{n} \psi_{\boldsymbol{\lambda}}} \ket{\partial_{m} \psi_{\boldsymbol{\lambda}}}  -  \bra{\partial_{n} \psi_{\boldsymbol{\lambda}}}\ket{\psi_{\boldsymbol{\lambda}}} \bra{\psi_{\boldsymbol{\lambda}}}\ket{\partial_{m} \psi_{\boldsymbol{\lambda}}} \right]}
\end{equation}
In the specific case of dipoles estimation, the symmetric logarithmic derivatives of a pure and non pure state can be written as \cite{chapeau2015optimized}
\begin{equation}
    \label{eq:SLD Bloch}
    \mathcal{L}_{n}=\left\{ \begin{array}{lr} - \frac{\vec{b} \partial_{n}\vec{b} }{1 - \vert \vec b_{\boldsymbol{\lambda}} \vert^2} \mathbb{I} + \left( \frac{\vec{b} \partial_{n}\vec{b} }{1 - \vert \vec b_{\boldsymbol{\lambda}} \vert^2} \vec{b} + \partial_{n}\vec{b} \right) \vec{\sigma} &  \mbox{if} \,\,\, \vert\vec b_{\boldsymbol{\lambda}} \vert \, < \, 1\\
    & \\
    \partial_{n}\vec{b} \cdot \vec{\sigma} & \mbox{if} \,\,\, \vert\vec b_{\boldsymbol{\lambda}} \vert \, = \, 1
    \end{array}\right.
\end{equation}
and the QFI matrix $\mathcal{I}(\boldsymbol{\lambda})_{n,m}$ can be expressed as \cite{zhong2013fisher,ragazzi2024generalized} 
\begin{equation}
\label{BlochQFI}
    \mathcal{I}(\boldsymbol{\lambda})_{n,m}=\left\{ \begin{array}{lr}
        (\partial_{n}\vec b_{\boldsymbol{\lambda}} \cdot \partial_{m}\vec b_{\boldsymbol{\lambda}}) +\frac{(\vec b_{\boldsymbol{\lambda}} \cdot \partial_{n}\vec b_{\boldsymbol{\lambda}} )(\vec b_{\boldsymbol{\lambda}} \cdot \partial_{m}\vec b_{\boldsymbol{\lambda}})}{1-\vert \vec b_{\boldsymbol{\lambda}} \vert^2} &  \mbox{if} \,\,\, \vert\vec b_{\boldsymbol{\lambda}} \vert \, < \, 1\\
        & \\
        (\partial_{n}\vec b_{\boldsymbol{\lambda}} \cdot \partial_{m}\vec b_{\boldsymbol{\lambda}}) & \mbox{if} \,\,\, \vert\vec b_{\boldsymbol{\lambda}} \vert \, = \, 1
    \end{array}\right. ,
\end{equation}
where $\vec b_{\boldsymbol{\lambda}}$ is the Bloch vector associated with the state $\rho_{\boldsymbol{\lambda}}$, and follows the definition of Eq.\eqref{eq:r}. In this framework, the Cramer-Rao bound itself is a relation between matrices; the variance matrix $\boldsymbol{V}$ and the inverse of the (Q)FI matrix, as
\begin{equation}
    \label{eq:Cov_matr_bound}
    \boldsymbol{V} \boldsymbol{(\lambda)} \geq \frac{1}{M} {\boldsymbol{\mathcal{F}}_{\Pi}}^{-1}\boldsymbol{(\lambda)} \geq \frac{1}{M} {\boldsymbol{\mathcal{I}}}^{-1}\boldsymbol{(\lambda)},
\end{equation}
where ${\boldsymbol{V}}_{nm}=\langle \lambda_{n} \lambda_{m}\rangle-\langle \lambda_{n} \rangle \langle \lambda_{m}\rangle$. 
Through a positive, real matrix (of dimension $dim(\boldsymbol{\lambda}) \cross dim(\boldsymbol{\lambda})$), the so-called weight matrix $\boldsymbol{W}$, it is possible to define scalar bounds. In this section we adopt as a first case $\boldsymbol{W}=\mathbb{I}$, while in Sec.\ref{sec:FuIm} we will use a different $\boldsymbol{W}$ to analyze the implications of the results of the present paper in subatomic multiparameter estimation. Considering $\boldsymbol{W}=\mathbb{I}$, we obtain the scalar bound
\begin{equation}
   \begin{split} 
    \label{eq:Var_lower_bound}
     &\hbox{Tr} \left[ {\boldsymbol{WV}} \right]  = \Tr \left[ {\boldsymbol{V}} \right] = \sum_{n} \hbox{Var}(\lambda_{n}) \geq \frac{1}{M} \Tr \left[ {\boldsymbol{W\mathcal{F}}_{\Pi}}^{-1}\boldsymbol{(\lambda)} \right] 
     \geq \frac{1}{M} \hbox{Tr} \left[ {\boldsymbol{W\mathcal{I}}}^{-1}\boldsymbol{(\lambda)} \right] = \frac{1}{M} \hbox{Tr} \left[ {\boldsymbol{\mathcal{I}}}^{-1}\boldsymbol{(\lambda)} \right]  \ ,
   \end{split} 
\end{equation}
i.e. a lower bound on the sum of the variances associated to the parameters contained in the vector $\boldsymbol{\lambda}$ in a joint estimation of two (or more) parameters. For the specific case of two unknown parameters (i.e. $\mu$ and $d$) the inverses can be easily evaluated, arriving at 

\begin{equation}
   \begin{split} 
    \label{eq:Var_lower_bound_2}
     &\Tr \left[ {\boldsymbol{V}} \right] \geq   \frac{ \Tr \left[ {\boldsymbol{\mathcal{F}}_{\Pi}} \boldsymbol{(\lambda)} \right] }{M \det[\boldsymbol{\mathcal{F}}_{\Pi} \boldsymbol{(\lambda)} ]}  \geq \frac{ \Tr \left[ {\boldsymbol{\mathcal{I}}} \boldsymbol{(\lambda)} \right] }{M \det[\boldsymbol{\mathcal{I}} \boldsymbol{(\lambda)} ]} \, .
   \end{split} 
\end{equation}
Eqs.\eqref{eq:Var_lower_bound}-\eqref{eq:Var_lower_bound_2} intrinsically require the (Q)FI matrix to be invertible, and so that $det(\boldsymbol{\mathcal{I}}(\boldsymbol{\lambda}))\neq0$. A metrological model compatible with multiparameter estimation is said to be \textit{non-sloppy}, and allows the joint estimation of two (or more) parameters under analysis \cite{gutenkunst2007universally,daniels2008sloppiness,cavazzoni2024characterization,adani2024critical}. Additionally, another fundamental tool in the quantum multi-parameter estimation is the so called Uhlmann curvature matrix, whose elements are defined as
\begin{equation}
    \label{eq:general_Uhlmann_matrix}
    {U}_{nm}\boldsymbol{(\lambda)} = -i \Tr\left[\rho_{\boldsymbol{\lambda}}\frac{\mathcal{L}_{n}\mathcal{L}_{m}-\mathcal{L}_{m}\mathcal{L}_{n}}{2}\right].
\end{equation}
${\boldsymbol{U}}_{nm}\boldsymbol{(\lambda)}\neq0$ indicates that there is an intrinsic additional quantum noise in the joint estimation of $\lambda_{n}$ and $\lambda_{m}$, meaning that a simultaneous estimation of more than one parameter is less precise that two (or more) separate single parameter estimation procedures. For pure states the matrix $\boldsymbol{U}$ can be expressed as
\begin{equation}
    \label{eq: Uhlmann Pure}
    {U}_{nm}\boldsymbol{(\lambda)} = 4 \Im{\left[ \bra{\partial_{n} \psi_{\boldsymbol{\lambda}}} \ket{\partial_{m} \psi_{\boldsymbol{\lambda}}}  -  \bra{\partial_{n} \psi_{\boldsymbol{\lambda}}}\ket{\psi_{\boldsymbol{\lambda}}} \bra{\psi_{\boldsymbol{\lambda}}}\ket{\partial_{m} \psi_{\boldsymbol{\lambda}}} \right]}
\end{equation}
and for the specific case of dipoles, using Eq.\eqref{eq:SLD Bloch}, $\boldsymbol{U}$ can be written as
\begin{equation}
     \label{eq: Uhlmann qubit }
       U_{nm}= \vec b_{\boldsymbol{\lambda}} \cdot \left( \partial_n \vec b_{\boldsymbol{\lambda}} \times \partial_m \vec b_{\boldsymbol{\lambda}} \right)
\end{equation}
where again $\vec b_{\boldsymbol{\lambda}}$ follows the definition of Eq. \eqref{eq:r}. Differently from the QFI case (see Eq.\ref{BlochQFI}) here the formula of the Uhlmann curvature matrix is the same for the pure and mixed states, and the difference among the two cases is only related to the Bloch vector's norm. Apart from the definition, this matrix introduces other scalar bounds in the multiparameter estimation \cite{sharma2025mitigating}, as it defines a chain of inequalities between the QFI and the FI matrix as
\begin{align}
    \label{eq:Bound Multiparameter}
      &\Tr \left[ {\boldsymbol{V}} \right] \geq \frac{\Tr \left[ {\boldsymbol{\mathcal{F}}_{\Pi}}^{-1}\boldsymbol{(\lambda)} \right]}{M} \nonumber \\
      &\geq (1+R) \frac{\Tr \left[ {\boldsymbol{\mathcal{I}}}^{-1}\boldsymbol{(\lambda)} \right]}{M} \geq \frac{\Tr \left[ {\boldsymbol{\mathcal{I}}}^{-1}\boldsymbol{(\lambda)} \right]}{M}
\end{align}
where the coefficient $R$ for a two parameter estimation procedure can be defined as
\begin{align}
    \label{eq:Bound Multiparameter R}
    R = \sqrt{\frac{\det(\boldsymbol{U})}{\det(\boldsymbol{\mathcal{I}})}} \, ,
\end{align}
with $0 \leq R \leq 1$ \cite{carollo2019quantumness}. This bound set the ultimate limit on how much information can be extracted in a joint estimation of the dipole moments $\mu$ and $d$ that define the EM properties of our system. Overall, the theory of single parameter estimation sets the ultimate limits of precision of a measurement in which only one of the dipole moments is unknown and all the other system parameters are set. Contrariwise, the multiparameter estimation theory is necessary when both the dipole moments are not known, and the intent of the measurement procedure is to simultaneously estimate both $\mu$ and $d$.

\section{Results and Discussion}
\label{sec:RaD}

The following metrological analysis of the system consists in the characterization of the magnetic and electric dipole moments of the system, i.e. $\mu$ and $d$, under the effect of the known components of the EM field $\vec{E}$ and $\vec{B}$. We examine two distinct settings: (i) where the initial quantum state of the system can be controlled and optimized to maximize the time evolution of the (quantum) Fisher information associated with the dipole parameters $\mu$ and $d$, and (ii) where the system is in a thermal state, representing a prototypical example of static non-ideal working conditions. 

\subsection{Dynamical Metrology: Pure Initial State}

In the first scenario we assume to have the control on the initial state of the system, such that a maximization of the QFI over all the possible $\ket{\psi(0)}$ is possible. In similar analysis \cite{jing2015maximal,candeloro2024dimension}, the ideal state that maximize the QFI for a parameter of a two level system was found to be in the form
\begin{equation}
    \label{eq:ideal_state}
    \ket{\psi_{opt}} = \frac{1}{\sqrt{2}} \left( \ket{\phi_{-}} + e^{i\eta} \ket{\phi_{+}} \right).
\end{equation}
This correspond to the state $\ket{\psi_{opt,z}}=\nicefrac{\left(\ket{0} + e^{i\eta} \ket{1}\right)}{\sqrt{2}}$ for a $z$ rotation and $\ket{\psi_{opt,x}}=\nicefrac{\left(\cos{\left( \frac{\alpha}{2} \right)}\ket{0} \pm i \sin{\left( \frac{\alpha}{2} \right)} \ket{1}\right)}{\sqrt{2}}$ for an $x$ rotation. In the present model, in which the Hamiltonian of Eq.\eqref{eq:EDM_Hamiltonian_red} is the sum of an $x$ and a $z$ rotation, both $\ket{\phi_{+}}$ and $\ket{\phi_{-}}$ depends on the dipole moments of the system, in general also their superposition does. Nonetheless by fine tuning the phase of Eq.\eqref{eq:ideal_state}, which can assume any value $\eta \in [0,2\pi)$, it is possible to obtain a general state that maximize the QFI of the system independently on both $\mu$ and $d$. Specifically, following the mathematical calculation of Appendix \ref{sec:AM2}, and by choosing $\eta = \pm \frac{\pi}{2}$ the superposition of $\ket{\phi_{+}}$ and $\ket{\phi_{-}}$ results to be
\begin{equation}
    \label{eq:ideal_state_indep}
    \ket{\psi_{opt}} = \frac{1}{\sqrt{2}} \left( \ket{\phi_{-}} + e^{\pm i \frac{\pi}{2}} \ket{\phi_{+}} \right) = \frac{1}{\sqrt{2}} \left( \ket{0} \pm i \ket{1} \right)
\end{equation}
which correspond to a state that maximize at the same time the QFI of an $x$ and a $z$ rotation \cite{cavazzoni2024coin,cavazzoni2024optimizing} without depending on any of the parameters of the system. Then, according to Eqs.\eqref{eq:EDM_Hamiltonian_red_parallel}-\eqref{eq:EDM_Hamiltonian_red_orthogonal} this is sufficient for the optimization of the QFI for both $\mu$ and $d$. This result can be applied in all the quantum systems currently adopted in experiments aimed to provide signatures of CP violations or, by extension, in all general procedures aimed to precisely characterize the electromagnetic properties of quantum systems with both electric and magnetic dipole moments.

\subsubsection{Parallel Dipole Moments Configuration}
As a first configuration we present the results valid for elementary particles, nucleons and nuclei in which there is only one privileged direction due to the symmetry of the system and the electric and magnetic moment are aligned in the same direction \footnote{We choose $z$ for simplicity}, i.e. we consider a system governed by the Hamiltonian presented in Eq.\eqref{eq:EDM_Hamiltonian_red_parallel}. 
\paragraph{Noiseless Time Evolution --} Since for the parallel dipole moments configuration the eigenvalues are $ \pm \omega_{\parallel}= \pm (\pm \mu \, B_z + d \, E_z)$ and the eigenvectors are $\{\ket{\phi_{-}},\ket{\phi_{+}}\}=\{\ket{0},\ket{1}\}$, the noiseless time evolution of the initial optimal state is
\begin{equation}
    \label{eq:time_evolution_optimal_parallel}
    \ket{\psi_{opt}^{\parallel}(t)} = \frac{1}{\sqrt{2}} \left( e^{i\omega_{\parallel} t}\ket{0} + e^{-i(\omega_{\parallel} t - \eta)}\ket{1} \right) \, .
\end{equation}
The initial phase $\eta$ is arbitrary since the time evolution consist only in a rotation around the $z$-axis. The QFI matrix, when the Hamiltonian of the system is Eq.\eqref{eq:EDM_Hamiltonian_red_parallel} and the initial state is Eq.\eqref{eq:time_evolution_optimal_parallel}, has elements
\begin{align}
    \label{eq:QFI parallel static fields}
     &\mathcal{I}^{\parallel,\psi}_{d,d} = 4 (E_z \cdot t)^2 \nonumber \\
     &\mathcal{I}^{\parallel,\psi}_{d,\mu} = 4 E_z \cdot (\pm B_z) \cdot t \nonumber \\
     &\mathcal{I}^{\parallel,\psi}_{\mu,\mu} = 4 (\pm B_z \cdot t)^2 \nonumber \\
     &\mathcal{I}^{\parallel,\psi}_{\mu,d} = 4 E_z \cdot (\pm B_z) \cdot t 
\end{align}
The diagonal elements (i.e. $\mathcal{I}^{\parallel,\psi}_{d,d}$ and $\mathcal{I}^{\parallel,\psi}_{\mu,\mu}$ ) by themselves  indicate that the precision achievable through a single parameter estimation measurement scales quadratically with time for both the electric and magnetic dipole moment, with a prefactor that depends on the intensity of the EM field applied. This means that in principle with a perfect control on the system, the precision of the measurement increases $\propto t^2$. The off diagonal elements (i.e. $\mathcal{I}^{\parallel,\psi}_{d,\mu}$ and $\mathcal{I}^{\parallel,\psi}_{\mu,d}$) are not related to the precision of single parameter estimation measurement, but define the value of the determinant of the QFI and consequently the sloppiness of the model, as
\begin{align}
    \label{eq:sloppiness parallel field configuration}
    det(\boldsymbol{\mathcal{I}}^{\parallel,\psi}) &= \mathcal{I}^{\parallel,\psi}_{d,d}  \cdot \mathcal{I}^{\parallel,\psi}_{\mu,\mu}  - \mathcal{I}^{\parallel,\psi}_{d,\mu}  \cdot \mathcal{I}^{\parallel,\psi}_{\mu,d}   \nonumber \\
    & = 0 \,\,\,\,\,\, \forall \, \mu,B_z,d,E_z,t.
\end{align}
Practically speaking this means that the properties of the system do not depend separately on the dipole moments, but rather on a linear combination of them. This is clearly evident from the eigenproblem solution of the model, since the eigenvectors are parameter independent and the eigenvalues of the system depends on a linear combination of the dipole moments weighted by the electric and magnetic field strengths, as $\omega_{\parallel}= (\pm \mu \, B_z + d \, E_z)$. The Uhlmann matrix is identically null, i.e.
\begin{equation}
    \label{eq:Uhlmann parallel}
    \boldsymbol{U}^{\parallel,\psi} = \mathbf{0} \,\,\,\,\,\, \forall \, \mu,B_z,d,E_z,t,
\end{equation}
meaning that there is no intrinsic quantum noise related to the non commutativity of the symmetric logarithmic derivative. Together with the considerations about the determinant of the QFI we can conclude that in the parallel dipole moments configuration the statistical model is compatible but sloppy, meaning that in every experiment that involves nucleons, nuclei, elementary particles or every quantum system with an Hamiltonian in the form of Eq.\eqref{eq:EDM_Hamiltonian_red_parallel} every measurement procedure will always provides informations related to a linear combination of $\mu$ and $d$. Then, the dipole moments in this configuration must be determined through separate measurement procedures in which all the other parameters that define the physics of the system are known, and the precision of the estimation is related to a specific diagonal element of the QFI matrix alone.
\paragraph{Noisy Time Evolution --} To extend the dynamical estimation to a scenario in which noise affect the quantum evolution we consider as initial state $\rho^{\parallel}(0) = \ket{\psi_{opt}^{\parallel}(0)}\bra{\psi_{opt}^{\parallel}(0)}$ and Eq.\eqref{eq:depolarized_evolution} as the equation governing the dynamic of the system, leading to an evolved state in the form
\begin{align}
    \label{eq:rho_optimal_parallel}
    &\rho^{\parallel}(t) = e^{-\gamma t} \left( \mathcal{U}_{r,\parallel}(t) \rho^{\parallel}(0) \mathcal{U}_{r,\parallel}^{\dagger}(t) \right) + \left( 1 - e^{-\gamma t} \right) \frac{\mathbb{I}}{2} \nonumber \\
    & = \left( 1 - e^{-\gamma t} \right) \frac{\mathbb{I}}{2} + e^{-\gamma t} \ket{\psi_{opt}^{\parallel}(t)}\bra{\psi_{opt}^{\parallel}(t)}
\end{align}
This dynamic results in a QFI which is strictly related to the pure case scenario through the matrix relation
\begin{equation}
    \label{eq:QFI parallel static fields noisy}
    \boldsymbol{\mathcal{I}}^{\parallel,\rho}  = e^{-2\gamma t} \boldsymbol{\mathcal{I}}^{\parallel,\psi}.
\end{equation}
Consequently it is no longer true that the QFI increases $\propto t^2$, but it exists an ideal time 
\begin{equation}
    \label{eq:ideal t QFI parallel depolarized}
    t^{\parallel,\rho}_{opt}=\frac{1}{\gamma}
\end{equation}
in which the QFI result maxima for both the electric and magnetic dipole moments, and assume values
\begin{equation} 
\label{eq:max QFI parallel depolarized}
 t=t^{\parallel,\rho}_{opt}:\begin{cases}
\begin{aligned}
    \max_{t} \mathcal{I}^{\parallel,\rho}_{d,d} = \frac{4 E_z^{2}}{\gamma^2 e^2}  \\
    \max_{t} \mathcal{I}^{\parallel,\rho}_{\mu,\mu} = \frac{4 B_z^{2}}{\gamma^2 e^2} 
\end{aligned}
\end{cases}
\end{equation}
while all the other relevant quantity of the statistical model remain unchanged, i.e. 
\begin{equation}
    \label{eq:sloppiness parallel field noisy}
    det(\boldsymbol{\mathcal{I}}^{\parallel,\rho})=det(\boldsymbol{\mathcal{I}}^{\parallel,\psi})=0 \,\,\, \forall \, \mu,B_z,d,E_z,t
\end{equation} 
and 
\begin{equation}
    \label{eq:Uhlmann Matrix parallel field noisy}
    \boldsymbol{U}^{\parallel,\rho} = \boldsymbol{U}^{\parallel,\psi} = 0 \,\,\, \forall \, \mu,B_z,d,E_z,t \,.
\end{equation} 
Then, the only effect of the noise is to reduce the maximal precision achievable in the estimation of the dipole moments, introducing an optimal estimation time, without affecting the other main characteristics of the parallel dipole moments statistical model.

\subsubsection{Orthogonal Dipole Moments Configuration}

In the second configuration we present results valid for systems in which the magnetic and electric dipole moments lie on orthogonal directions, such as in heavy polar molecules \footnote{Here without loss of generality we choose $x$ and $z$ as directions}, i.e. we consider a system governed by the Hamiltonian presented in Eq.\eqref{eq:EDM_Hamiltonian_red_orthogonal}. 
\paragraph{Noiseless Time Evolution --} In the orthogonal dipole moments configuration $ \omega_{\perp} = \sqrt{(\mu B_{x})^2 + (d E_{z})^2}$ and $\{\ket{\phi_{-}},\ket{\phi_{+}}\} \neq \{\ket{0},\ket{1}\}$, then, the noiseless time evolution of a general initial optimal state is
\begin{equation}
    \label{eq:time_evolution_optimal_orthogonal}
    \ket{\psi_{opt}^{\perp}(t)} = \frac{1}{\sqrt{2}} \left( e^{i\omega_{\perp} t} \ket{\phi_{-}} + e^{-i(\omega_{\perp} t - \eta)} \ket{\phi_{+}} \right)
\end{equation}
For practical purposes, according to Eq.\eqref{eq:ideal_state_indep}, we can analyze the dynamic of
\begin{equation}
    \label{eq:time_evolution_optimal_orthogonal_indep}
    \ket{\psi_{opt}^{\perp}(t)} = \mathcal{U}_{r}(t) \left[  \frac{1}{\sqrt{2}} \left( \ket{0} \pm i \ket{1} \right) \right] ,
\end{equation}
since it inherits all the properties of the general optimal state of Eq.\eqref{eq:time_evolution_optimal_orthogonal} with the initial state being independent of all the parameters that define the system. The elements of the QFI matrix for the estimation of the dipole moments, when the Hamiltonian of the system is Eq.\eqref{eq:EDM_Hamiltonian_red_parallel} and the initial state is Eq.\eqref{eq:time_evolution_optimal_orthogonal_indep}, are
\begin{align}
    \label{eq:QFI orthogonal static fields diagonal elements}
    & \mathcal{I}^{\perp,\psi}_{d,d} = E_{z}^2 \left[ 4\frac{\zeta^2 }{\omega_{\perp}^2} t^2 + \frac{\chi^2}{\omega_{\perp}^4} \sin^2{\left( 2 \omega_{\perp} t \right)} \right] \nonumber \\
    & \mathcal{I}^{\perp,\psi}_{d,\mu} = ( \chi \zeta B_x E_z ) \left[ 4\frac{t^2}{\omega_{\perp}^2}  - \frac{\sin^2{\left( 2 \omega_{\perp} t \right)}}{\omega_{\perp}^4}  \right] \nonumber \\
    & \mathcal{I}^{\perp,\psi}_{\mu,\mu} = B_{x}^2 \left[ 4\frac{\chi^2}{\omega_{\perp}^2} t^2 + \frac{\zeta^2}{\omega_{\perp}^4} \sin^2{\left(2 \omega_{\perp} t \right)} \right] \nonumber \\
    & \mathcal{I}^{\perp,\psi}_{\mu,d} = ( \chi \zeta B_x E_z ) \left[ 4\frac{t^2}{\omega_{\perp}^2}  - \frac{\sin^2{\left( 2 \omega_{\perp} t \right)}}{\omega_{\perp}^4}  \right] \, ,
\end{align}
where here $\chi = \mu \cdot B_{x}$ and $\zeta = d \cdot E_{z}$. The diagonal elements (i.e. $\mathcal{I}^{\perp,\psi}_{d,d}$ and $\mathcal{I}^{\perp,\psi}_{\mu,\mu}$ ) indicate that the precision achievable through a measurement has a quadratic time dependence $\propto t^2$, with an additive oscillating $\sin^2{\left( 2 \omega_{\perp} t \right)}$ term for both the electric and magnetic dipole moments. The off diagonal elements (i.e. $\mathcal{I}^{\perp,\psi}_{d,\mu}$ and $\mathcal{I}^{\perp,\psi}_{\mu,d}$) are related to the precision of the joint estimation of $\mu$ and $d$, since they define the value of the determinant, which reads
\begin{align}
    \label{eq:sloppiness orthogonal field configuration}
    det(\boldsymbol{\mathcal{I}}^{\perp,\psi}) &= \mathcal{I}^{\perp,\psi}_{d,d}  \cdot \mathcal{I}^{\perp,\psi}_{\mu,\mu}  - \mathcal{I}^{\perp,\psi}_{d,\mu}  \cdot \mathcal{I}^{\perp,\psi}_{\mu,d}  \nonumber \\
    & = \frac{4(B_x E_z t)^2\sin^2{\left( 2 \omega_{\perp} t \right)}}{\omega^{2}_{\perp}}
\end{align}
The determinant of the QFI matrix is then always non-null except for the zeros of $\sin^2{\left( 2 \omega_{\perp} t \right)}$. Practically speaking, the properties of the system does depend separately on the electric and magnetic dipole moments, for all t except for $t^{*}=\nicefrac{n\pi}{2 \omega_{\perp}}$ with $n \in \mathbb{N}$. This means that the system sloppiness is time dependent, and the statistical model depends on a linear combination of the dipole moments of the system for $t^{*}$, while during all the other time the system is non-sloppy. This condition is in contrast with the parallel dipole moments configuration in which the sloppiness of the statistical model was valid for every $t$ (see Eq.\eqref{eq:sloppiness parallel field configuration}). The Uhlmann matrix has elements
\begin{align}
    \label{eq:Uhlmann orthogonal}
    & U^{\perp,\psi}_{d,d} = 0 \nonumber \\
    & U^{\perp,\psi}_{d,\mu} = + \frac{2 B_x E_z t \sin{(2 \omega_{\perp} t)}}{\omega_{\perp}}  \nonumber \\
    & U^{\perp,\psi}_{\mu,\mu} = 0 \nonumber \\
    & U^{\perp,\psi}_{\mu,d} = - \frac{2 B_x E_z t \sin{(2 \omega_{\perp} t)}}{\omega_{\perp}}
\end{align}
meaning that there is an intrinsic quantum noise related to the non commutativity of the symmetric logarithmic derivative that leads to a parameter $R^{\perp,\psi} = \sqrt{\nicefrac{\det(\boldsymbol{U}^{\perp,\psi})}{\det(\boldsymbol{\mathcal{I}^{\perp,\psi}})}} = 1$, i.e. maximally incompatible parameters \cite{razavian2020quantumness}. Together with the considerations about the determinant of the QFI we can conclude that in the orthogonal dipole moments configuration the statistical model is not compatible and sloppy only at specific times $t^{*}$. Then, differently from the parallel dipole moments configuration it is possible to jointly estimate both the dipole moments $\mu$ and $d$ of a quantum system for almost all the $t$ except for $t^{*}$.
\paragraph{Noisy Time Evolution --} Including the potential effect of a noisy evolution we consider $\rho^{\perp}(0) = \ket{\psi_{opt}^{\perp}(0)}\bra{\psi_{opt}^{\perp}(0)}$ as initial state and Eq.\eqref{eq:depolarized_evolution} as the Lindblad evolution, leading to
\begin{align}
    \label{eq:rho_optimal_orthogonal}
    &\rho^{\perp}(t) = e^{-\gamma t} \left( \mathcal{U}_{r,\perp}(t) \rho^{\perp}(0) \mathcal{U}_{r,\perp}^{\dagger}(t) \right) + \left( 1 - e^{-\gamma t} \right) \frac{\mathbb{I}}{2} \nonumber \\ 
    & = \left( 1 - e^{-\gamma t} \right) \frac{\mathbb{I}}{2} + e^{-\gamma t} \ket{\psi_{opt}^{\perp}(t)}\bra{\psi_{opt}^{\perp}(t)}
\end{align}
This dynamic leads to a QFI which is strictly related to the pure case scenario through the matrix relation
\begin{equation}
    \label{eq:QFI orthogonal static fields noisy}
    \boldsymbol{\mathcal{I}}^{\perp,\rho}  = e^{-2\gamma t} \boldsymbol{\mathcal{I}}^{\perp,\psi}.
\end{equation}
Consequently it is no longer true that the precision related to the estimation of the electric and magnetic dipole moments increases in time with a leading $t^2$ component and an additional oscillating $\sin^2{(2\omega t)}$ term. Indeed, the exponentially decaying term introduces an optimal time in which the diagonal elements of the QFI result maximized. Since the QFI is the sum of two terms we distinguish two cases, the first scenario in which $\zeta \gg \chi$ and the second one where $\zeta \ll \chi$. Even if in many systems one of the two dipole components result higher (i.e. in the electron or neutron), since the dynamic of the depends on $\chi=\mu \cdot B_{x}$ and $\zeta = d \cdot E_{z}$ both regimes are of metrological interest, since with the tuning of the magnetic and electric field it is possible to study both $\zeta \gg \chi$ or $\zeta \ll \chi$. In the first case, the maximal values of the QFI are
\begin{equation} 
\label{eq:max QFI orthogonal depolarized zeta dominant}
\zeta \gg \chi : \begin{cases}
\begin{aligned}
    \max_{t} \mathcal{I}^{\perp,\rho}_{d,d} &= 4\frac{E_{z}^2 \zeta^2 }{\omega_{\perp}^2 e^2 \gamma^2} && \text{for } t=\frac{1}{\gamma} \\
    \max_{t} \mathcal{I}^{\perp,\rho}_{\mu,\mu} &= 4 E_{z}^{2}\frac{\zeta^{2} e^{\left[-\frac{\gamma}{\omega_{\perp}}\arctan\!\left(\frac{2\omega_{\perp}}{\gamma}\right)\right]}}{\omega_{\perp}^{2}\left(\gamma^{2}+4\omega_{\perp}^{2}\right)} && \text{for } t=\frac{1}{2 \omega_{\perp}} \arctan\left( \frac{2 \omega_{\perp}}{\gamma} \right)
\end{aligned}
\end{cases}
\end{equation}
In the second case, the maximal values of the QFI are
\begin{equation} 
\label{eq:max QFI orthogonal depolarized chi dominant}
\zeta \ll \chi : \begin{cases}
\begin{aligned}
    \max_{t} \mathcal{I}^{\perp,\rho}_{d,d} &= 4 B_{x}^{2}\frac{\chi^{2} e^{\left[-\frac{\gamma}{\omega_{\perp}}\arctan\!\left(\frac{2\omega_{\perp}}{\gamma}\right)\right]}}{\omega_{\perp}^{2}\left(\gamma^{2}+4\omega_{\perp}^{2}\right)} && \text{for } t=\frac{1}{2 \omega_{\perp}} \arctan\left( \frac{2 \omega_{\perp}}{\gamma} \right) \\
    \max_{t} \mathcal{I}^{\perp,\rho}_{\mu,\mu} &= 4\frac{B_{x}^2 \chi^2 }{\omega_{\perp}^2 e^2 \gamma^2} && \text{for } t=\frac{1}{\gamma}
\end{aligned}
\end{cases}
\end{equation}
Then, for the orthogonal configuration the maximum of the QFI of the electric and magnetic dipole moments are in general reached at different times. Indeed, depending on the specific dipole moment that we want to estimate with an higher precision the measure should be done at a different specific time. The other relevant quantities of the statistical model, differently from the parallel dipole moments scenario, result affected by the noise, i.e. 
\begin{equation}
    \label{eq:sloppiness orthogonal field noisy}
    det(\boldsymbol{\mathcal{I}}^{\perp,\rho})= e^{-4\gamma t} det(\boldsymbol{\mathcal{I}}^{\perp,\psi}) \, ,
\end{equation} 
and 
\begin{equation}
    \label{eq:Uhlmann Matrix orthogonal field noisy}
    \boldsymbol{U}^{\perp,\rho} = e^{-3\gamma t} \boldsymbol{U}^{\perp,\psi} \, .
\end{equation} 
This means that the main characteristic of the system are preserved for small $t$, as $R^{\perp,\rho}= e^{-\gamma t} R^{\perp,\psi} = e^{-\gamma t}$; however, the noise leads to a statistical model in which the quantumness of the multiparameter estimation exponentially vanishes over time. Considering all the characteristics of the physical models, the orthogonal dipole moments configuration results metrologically more relevant in the joint multiparameter estimation of the dipole moments since it almost always non-sloppy.

\subsection{Static Metrology: Thermal Equilibrium State}

After the complete characterization of the system in the dynamical framework in both ideal and noisy scenario it is important to study how the thermalization of the system impact its metrological characterization \cite{daniotti2018qubit}. In this condition an optimization is no longer possible, since every thermal quantum system is defined by the Gibbs state presented in Eq.\eqref{eq:thermal_rho}. The elements of the energy FI matrix of such configurational structure can be obtained directly from Eq.\eqref{eq:fisher_information} as
\begin{align}
\label{eq:fisher_information_th_2}
(\mathcal{F}_{\mathcal{E}})_{n,m} & = \dfrac{ (\partial_{n} \rho \left( \mathcal{E}_+ \vert \boldsymbol{\lambda} \right)) (\partial_{m} \rho \left( \mathcal{E}_+ \vert \boldsymbol{\lambda} \right)) }{\rho\left( \mathcal{E}_+ \vert d \right)} \nonumber \\
& + \dfrac{ (\partial_{n} \rho \left( \mathcal{E}_- \vert \boldsymbol{\lambda} \right)) (\partial_{m} \rho \left( \mathcal{E}_- \vert \boldsymbol{\lambda} \right)) }{\rho\left( \mathcal{E}_- \vert d \right)} \nonumber \\
& = \dfrac{ (\partial_{n} \rho \left( \mathcal{E}_+ \vert \boldsymbol{\lambda} \right)) (\partial_{m} \rho \left( \mathcal{E}_+ \vert \boldsymbol{\lambda} \right)) }{\rho\left( \mathcal{E}_+ \vert d \right) (1-\rho\left( \mathcal{E}_+ \vert d \right))} \nonumber \\
& = \dfrac{ (\partial_{n} \rho \left( \mathcal{E}_- \vert \boldsymbol{\lambda} \right)) (\partial_{m} \rho \left( \mathcal{E}_- \vert \boldsymbol{\lambda} \right)) }{\rho\left( \mathcal{E}_- \vert d \right) (1-\rho\left( \mathcal{E}_- \vert d \right))},
\end{align}
where the energy measurement is performed through the measure of the population of the eigenstates $\Pi_{\pm}=\{ \ket{\phi_{\pm}}\bra{\phi_{\pm}}\}$. The QFI results in the sum of the FI to an additional term as
\begin{align}
\label{eq:quantum_fisher_information_th}
(\mathcal{I})_{n,m} &= (\mathcal{F}_{\mathcal{E}})_{n,m} + \tanh^2{(\beta \omega)} \cdot \bigg( \bra{\phi_+}\ket{\partial_{n} \phi_-} \bra{\partial_{m} \phi_-}\ket{\phi_+} \nonumber \\
&+ \bra{\phi_-}\ket{\partial_n \phi_+} \bra{\partial_m \phi_+}\ket{\phi_-}  \bigg) \nonumber \\
\end{align}

\subsubsection{Parallel Dipole Moments Configuration}

In the parallel dipole moments configuration (i.e. when $\omega=\omega_{\parallel}=(\pm \mu \, B_z + d \, E_z)$ and the eigenvectors are $\{\ket{\phi_{-}},\ket{\phi_{+}}\}=\{\ket{0},\ket{1}\}$) the quantum Fisher information for a thermal state $\rho^{\parallel}_{th}$ coincides with the classical Fisher information of an energy measurement,
\begin{equation}
    \label{eq:(Q)FI thermal parallel}
    \boldsymbol{\mathcal{I}}^{\parallel,\rho_{th}}=\boldsymbol{\mathcal{F}}^{\parallel,\rho_{th}}_{\mathcal{E}} \,.
\end{equation}
This directly derive from Eq.\eqref{eq:quantum_fisher_information_th}, since the additional term present is identically null for all the possible values of the parameters that describe the system, as the eigenvectors of the system are independent of $\mu,d$. This means that the study of the population of the energy levels correspond to the ideal metrological scheme for the estimation of the dipole moments $\mu$ and $d$. Particularly, for the specific system in analysis Eq.\eqref{eq:fisher_information_th_2} reduces to
\begin{align}
\label{eq:fisher_information_th_result_parallel}
&\mathcal{I}^{\parallel,\rho_{th}}_{d,d}=(\mathcal{F}^{\parallel,\rho_{th}}_{\mathcal{E}})_{d,d} = \beta^2 E_{z}^{2}  \sech^2{(\beta \omega_{\parallel})} \nonumber \\
&\mathcal{I}^{\parallel,\rho_{th}}_{d,\mu}=(\mathcal{F}^{\parallel,\rho_{th}}_{\mathcal{E}})_{d,\mu} = \beta^2 (\pm B_{z}) \cdot E_{z} \sech^2{(\beta \omega_{\parallel})}  \nonumber \\
&\mathcal{I}^{\parallel,\rho_{th}}_{\mu,\mu}=(\mathcal{F}^{\parallel,\rho_{th}}_{\mathcal{E}})_{\mu,\mu} = \beta^2 B_{z}^{2} \sech^2{(\beta \omega_{\parallel})}  \nonumber \\
&\mathcal{I}^{\parallel,\rho_{th}}_{\mu,d}=(\mathcal{F}^{\parallel,\rho_{th}}_{\mathcal{E}})_{\mu,d} =  \beta^2 (\pm B_{z}) \cdot E_{z} \sech^2{(\beta \omega_{\parallel})} 
\end{align}
Even if for a thermal state an optimization procedure over the state cannot be performed, there exist an ideal temperature for which the estimation of both the magnetic and electric dipole moments results optimal. This specific temperature, simultaneously maximize the diagonal elements of the $\boldsymbol{\mathcal{F}}^{\parallel,\rho_{th}}_{\mathcal{E}}$ matrix, and corresponds to the temperature that satisfies the equation $\beta \omega_{\parallel} \tanh{(\beta \omega_{\parallel})} = 1$, leading to
\begin{equation}
    \label{eq:max FI thermal parallel}
     T^{\parallel}_{opt,\mathcal{F}_{\mathcal{E}}} \approx \frac{\omega_{\parallel}}{1.1997 \cdot \kappa_{B}} = \frac{(\pm \mu \, B_z + d \, E_z)}{1.1997 \cdot \kappa_{B}} \, ,
\end{equation}
which depends on the specific values of $\mu,d,B_z,E_z$ and then changes for every experimental setup. This temperature lead to a maximal value of the diagonal elements of the (Q)FI in the form
\begin{equation} 
\label{eq:max QFI parallel thermal}
 T=T^{\parallel}_{opt,\mathcal{F}_{\mathcal{E}}}:\begin{cases}
\begin{aligned}
    \max_{T} \mathcal{I}^{\parallel,\rho_{th}}_{d,d} = \max_{T} (\mathcal{F}_{\mathcal{E}}^{\parallel,\rho_{th}})_{d,d} = 0.4392 E_{z}^{2} \\
    \max_{T} \mathcal{I}^{\parallel,\rho_{th}}_{\mu,\mu} = \max_{T} (\mathcal{F}_{\mathcal{E}}^{\parallel,\rho_{th}})_{\mu,\mu} = 0.4392 B_{z}^{2}
\end{aligned}
\end{cases}
\end{equation}
The off diagonal elements of $\boldsymbol{\mathcal{F}}^{\parallel,\rho_{th}}_{\mathcal{E}}$ are again not related to the precision of the measurement, but can be again exploited to calculate the value of the determinant and consequently the sloppiness of the model for every temperature, as
\begin{align}
    \label{eq:sloppiness parallel field configuration thermal state}
    det(\boldsymbol{\mathcal{I}}^{\parallel,\rho_{th}}) &= \mathcal{I}^{\parallel,\rho_{th}}_{d,d} \cdot \mathcal{I}^{\parallel,\rho_{th}}_{\mu,\mu}  - \mathcal{I}^{\parallel,\rho_{th}}_{d,\mu}  \cdot \mathcal{I}^{\parallel,\rho_{th}}_{\mu,d} \nonumber \\
    & = 0 \,\,\,\,\,\, \forall \, \mu,B_z,d,E_z,T \,.
\end{align}
Then, also for the thermal configuration the statistical model for the parallel dipole moments configuration results completely sloppy independently of all the parameters that define the system itself. On the other hand, as for the dynamical case the Uhlmann matrix of a thermal state in the parallel dipole moments configuration results null, as
\begin{equation}
    \label{eq:Uhlmann Matrix parallel field thermal}
    \boldsymbol{U}^{\parallel,\rho_{th}} = 0 \,\,\, \forall \, \mu,B_z,d,E_z,T \,.
\end{equation} 
We can then conclude that the parallel dipole moments configuration in both static and dynamic conditions is a valid resource only for the separate estimation of electric and magnetic dipole moments since our analysis illustrate that in both static and dynamic conditions the parallel dipole moments statistical model results compatible but sloppy.

\subsubsection{Orthogonal Dipole Moments Configuration}

In the orthogonal dipole moments configuration (i.e. when $\omega=\omega_{\perp}=\sqrt{(\mu B_{x})^2 + (d E_{z})^2}$ and the eigenvectors are defined by Eq.\eqref{eq:eigenstate_n3cell}) the FI of a thermal state $\rho_{th}^{\perp}$ is
\begin{align}
\label{eq:fisher_information_th_result_orthogonal}
&(\mathcal{F}^{\perp,\rho_{th}}_{\mathcal{E}})_{d,d} = \frac{\beta^2 E_{z}^{2} \zeta^2 \sech^2{(\beta \omega_{\perp})}}{\omega_{\perp}^4} \nonumber \\
&(\mathcal{F}^{\perp,\rho_{th}}_{\mathcal{E}})_{d,\mu} = \frac{\beta^2 E_{z} \zeta B_{x} \chi \sech^2{(\beta \omega_{\perp})}}{\omega_{\perp}^4}  \nonumber \\
&(\mathcal{F}^{\perp,\rho_{th}}_{\mathcal{E}})_{\mu,\mu} = \frac{\beta^2 B_{x}^{2} \chi^2 \sech^2{(\beta \omega_{\perp})}}{\omega_{\perp}^4} \nonumber \\
&(\mathcal{F}^{\perp,\rho_{th}}_{\mathcal{E}})_{\mu,d} = \frac{\beta^2 E_{z} \zeta B_{x} \sech^2{(\beta \omega_{\perp})}}{\omega_{\perp}^4}
\end{align}
As for the parallel dipole moments configuration, also in the orthogonal dipole moments configuration there exist an ideal temperature for which the estimation of both the magnetic and electric dipole moments results optimal, maximizing both the diagonal elements of $\boldsymbol{\mathcal{F}}^{\perp,\rho_{th}}_{\mathcal{E}}$. In this specific case, the ideal temperature satisfies the almost identical equation $\beta \omega_{\perp} \tanh{(\beta \omega_{\perp})} = 1$, leading to
\begin{equation}
    \label{eq:max FI thermal orthogonal}
     T^{\perp}_{opt,\mathcal{F}_{\mathcal{E}}} \approx \frac{\omega_{\perp}}{1.1997 \cdot \kappa_{B}} = \frac{\sqrt{(\mu B_{x})^2 + (d E_{z})^2}}{1.1997 \cdot \kappa_{B}} \, .
\end{equation}
This temperature lead to a maximal value of the diagonal elements of the (Q)FI in the form
\begin{equation} 
\label{eq:max QFI orthogonal thermal}
 T=T^{\perp}_{opt,\mathcal{F}_{\mathcal{E}}}:\begin{cases}
\begin{aligned}
    \max_{T} (\mathcal{F}_{\mathcal{E}}^{\perp,\rho_{th}})_{d,d} = 0.4392  \frac{ E_{z}^{2} \zeta^2 }{\omega_{\perp}^4} \nonumber \\
    \max_{T} (\mathcal{F}_{\mathcal{E}}^{\perp,\rho_{th}})_{\mu,\mu} = 0.4392 \frac{ B_{x}^{2} \chi^2 }{\omega_{\perp}^4}
\end{aligned}
\end{cases}
\end{equation}
The analysis of the population of the energy levels in the orthogonal dipole moments configuration is then a valid metrological resource. Nonetheless, differently from the dynamical procedure, having orthogonal dipole moments results in a compatible but sloppy model, since
\begin{align}
    \label{eq:sloppiness orthogonal field configuration thermal state}
    det(\boldsymbol{{\mathcal{F}}}_{\mathcal{E}}^{\perp,\rho_{th}}) &= (\mathcal{F}^{\perp,\rho_{th}}_{\mathcal{E}})_{d,d} \cdot (\mathcal{F}^{\perp,\rho_{th}}_{\mathcal{E}})_{\mu,\mu}  \nonumber \\
    & - (\mathcal{F}^{\perp,\rho_{th}}_{\mathcal{E}})_{d,\mu}  \cdot (\mathcal{F}^{\perp,\rho_{th}}_{\mathcal{E}})_{\mu,d} \nonumber \\
    & = 0 \,\,\,\,\,\, \forall \, \mu,B_x,d,E_z,T \, ,
\end{align}
and
\begin{equation}
    \label{eq:Uhlmann Matrix orthogonal field thermal energy}
    \boldsymbol{U}^{\perp,\rho_{th}} = 0 \,\,\, \forall \, \mu,B_z,d,E_z,T \,.
\end{equation} 
In this condition the FI does not equal the QFI of the model as the contribution that arise from the derivative of the eigenvectors is non-null and lead to the definition of the matrix elements $(\mathcal{I}^{\perp,\rho_{th}})_{n,m}$ in the form
\begin{align}
\label{eq:quantum_fisher_information_th_result_orthogonal}
&(\mathcal{I}^{\perp,\rho_{th}})_{d,d} = (\mathcal{F}^{\perp,\rho_{th}}_{\mathcal{E}})_{d,d} + \frac{ E_{z}^2 \chi^2 \tanh^2\left(\beta \omega_{\perp} \right)}{2\omega_{\perp}^{4}} \nonumber \\
&(\mathcal{I}^{\perp,\rho_{th}})_{d,\mu} = (\mathcal{F}^{\perp,\rho_{th}}_{\mathcal{E}})_{d,\mu} - \frac{ E_{z} \zeta B_{x} \chi \tanh^2\left(\beta \omega_{\perp} \right)}{2\omega_{\perp}^{4}} \nonumber \\
&(\mathcal{I}^{\perp,\rho_{th}})_{\mu,\mu} = (\mathcal{F}^{\perp,\rho_{th}}_{\mathcal{E}})_{\mu,\mu} + \frac{ B_{x}^2 \zeta^2 \tanh^2\left(\beta \omega_{\perp} \right)}{2\omega_{\perp}^{4}}  \nonumber \\
&(\mathcal{I}^{\perp,\rho_{th}})_{\mu,d} = (\mathcal{F}^{\perp,\rho_{th}}_{\mathcal{E}})_{\mu,d} - \frac{ E_{z} \zeta B_{x} \chi \tanh^2\left(\beta \omega_{\perp} \right)}{2\omega_{\perp}^{4}}
\end{align}
As the QFI is the sum of the FI matrix of an energy measurement plus an additional term with a different temperature dependence, the position of the maximum is different with respect to Eq.\eqref{eq:max FI thermal orthogonal}, and depends on the relative value of $\chi$ and $\zeta$. Indeed for $\zeta \gg \chi$
\begin{equation}
\label{eq:max QFI orthogonal thermal zeta dominant}
\zeta \gg \chi : \begin{cases}
\begin{aligned}
    &\max_{T} \mathcal{I}^{\perp,\rho_{th}}_{d,d} = 0.4392 \frac{ E_{z}^2 \chi^2 }{\omega_{\perp}^{4}} && \text{for } T = T^{\perp}_{opt,\mathcal{F}_{\mathcal{E}}}   \\
    &\max_{T} \mathcal{I}^{\perp,\rho_{th}}_{\mu,\mu} = \frac{ B_{x}^2 \zeta^2}{2\omega_{\perp}^{4}} && \text{for } T \rightarrow 0
\end{aligned}
\end{cases}
\end{equation}
while for $\zeta \ll \chi$
\begin{equation}
\label{eq:max QFI orthogonal thermal chi dominant}
\zeta \ll \chi : \begin{cases}
\begin{aligned}
    &\max_{T} \mathcal{I}^{\perp,\rho_{th}}_{d,d} =  \frac{ E_{z}^2 \chi^2}{2\omega_{\perp}^{4}}  && \text{for } T \rightarrow 0 \\
    &\max_{T} \mathcal{I}^{\perp,\rho_{th}}_{\mu,\mu} = 0.4392 \frac{  B_{x}^2 \zeta^2}{2\omega_{\perp}^{4}}  && \text{for } T \approx T^{\perp}_{opt,\mathcal{F}_{\mathcal{E}}} 
\end{aligned}
\end{cases}
\end{equation}
Meaning that the ideal temperature to reach the maximum of the diagonal elements of the QFI in the thermal orthogonal dipole moments configuration depends on the parameters of the system and it is different for the electric and magnetic dipole moments. For the other combination of $\chi$ and $\zeta$ the maximum changes accordingly to Eq.\eqref{eq:quantum_fisher_information_th_result_orthogonal}. Following the results of Eqs.\eqref{eq:max QFI orthogonal thermal zeta dominant}-\eqref{eq:max QFI orthogonal thermal chi dominant} a clarification has to be done. Physically it is impossible to reach $T=0$, and so the maximum value of the QFI depends on the lowest temperature achievable in the experimental set-up adopted. Apart from this consideration, it is possible to calculate the determinant of the QFI matrix at all temperatures as
\begin{align}
    \label{eq:quantum sloppiness orthogonal field configuration thermal state}
    &det(\boldsymbol{\mathcal{I}}^{\perp,\rho_{th}}) = \mathcal{I}^{\perp,\rho_{th}}_{d,d} \cdot \mathcal{I}^{\perp,\rho_{th}}_{\mu,\mu}  - \mathcal{I}^{\perp,\rho_{th}}_{d,\mu}  \cdot \mathcal{I}^{\perp,\rho_{th}}_{\mu,d} \nonumber \\
    & = \frac{B_{x}^2 E_{z}^2 \beta^2 \sech^2\left(\beta \omega_{\perp} \right) \tanh^2\left(\beta \omega_{\perp}\right)}{2 \omega_{\perp}^2}.
\end{align}
and it is never zero for non trivial configurations (i.e. it is zero only when at least one of the two fields $E_{z}$ or $B_{x}$ is zero). This means that the QFI is always non singular and the sloppiness of the statistical model observed in Eq.\eqref{eq:sloppiness orthogonal field configuration thermal state} was due  to the choice of the energy measurement. Additionally, the Uhlmann matrix of a thermal state in the orthogonal dipole moments configuration vanishes
\begin{equation}
    \label{eq:Uhlmann Matrix orthogonal field thermal}
    \boldsymbol{U}^{\perp,\rho_{th}} = 0 \,\,\, \forall \, \mu,B_z,d,E_z,T \,.
\end{equation} 
Differently from its dynamical counterpart, in the thermal configuration the orthogonal dipole moments statistical model is both compatible and non sloppy, resulting a better resource for a joint estimation of the electric and magnetic dipole moments compared to the parallel dipole moments configuration. 

\section{Fundamental Implications}
\label{sec:FuIm}

The results of Sec.\ref{sec:RaD} intrinsically implies theoretical limits for the measurement of fundamental properties of elementary or subatomic particles such as the electron or the neutron. As mentioned in Sec.\ref{sec:I}, a measurable neutron EDM would be clear evidence of CP violation and is essential for the explanation of the matter-antimatter asymmetry in the universe, indicating also new sources of CP violation beyond the standard model possibly related to baryogenesis. Additionally, in the Standard Model, the electron EDM arises from the CP-violating components of the CKM matrix and can be the experimental evidence of the accuracy of technicolor models \cite{pospelov2005electric}. The precise characterization of the neutron and electron dipole moments is then fundamental in high energy physics experiments. We followingly describe the necessary conditions to obtain reliable results in experiments aimed to characterize the electron or neutron dipole moments in physical systems.

\subsection{Neutron Electric Dipole Moment}

The neutron has a magnetic moment $\mu_{n}$, and, as mentioned in Sec.\ref{sec:I}, despite being electrically neutral, may have an electric dipole moment, linked to the separation of positive and negative charge along the spin axis, contemporary showing a magnetic moment because of its quark composition $(udd)$ \cite{pospelov2001neutron}.  In experiments aimed to measure the dipole moment of the neutron \cite{baker2006improved} the physical model correspond to the parallel dipole moments configuration of the present paper, as they are based on the measurement of the neutron Larmor angular velocity in an electromagnetic field, i.e.
\begin{equation}
    \label{eq:larmor neutron}
    \omega_{n} = \pm \mu \cdot B_{z} + d_{n} \cdot E_{z}  \, .
\end{equation}
The sloppiness of the model implies that every experimental procedure aimed to characterize the dipolar properties of the neutron needs a precise prior knowledge of either the magnetic or electric dipole moment. Additionally, the statistical model implies that the variance about the measurement of $\omega_{n}$ is the sum of the variance of $\Delta( \mu_{n} \cdot B_{z})$ and $\Delta( d_{n} \cdot E_{z})$, and so the control on the EM field applied should be sufficient to state that $\nicefrac{\Delta(d_{n})}{d_{n}} \ll 1$, i.e.
\begin{equation}
    \label{eq:condition neutron}
    \mathsmaller{ \sqrt{  \left( \frac{ \Delta(\omega_{n}) + ( \mu \cdot \Delta(B_{z}) + B_{z} \cdot \Delta(\mu)) }{ \omega_{n} + \mu \cdot B_{z} } \right)^2 + \left( \frac{\Delta( E_{z})}{E_{z}} \right)^2  } \ll 1 }
\end{equation}
in order to have a physically meaningful result. Using the definition of the (Q)FI this means that the number of measurements required to obtain reliable results in the estimation of the value of the magnetic dipole moment of the neutron have to satisfy the relation
\begin{equation}
    \label{eq:measurement neutron}
    M^{\parallel}_{n} \gg \frac{1}{ d_{n}^2 \cdot \mathcal{I}^{\parallel}_{d_{n},d_{n}} } \, ,
\end{equation}
where $\mathcal{I}^{\parallel}_{d_{n},d_{n}}$ can assume the values illustrated in Sec.\ref{sec:RaD} depending on the nature of the estimation procedure itself. For a precise dynamical procedure  of $\mathcal{I}^{\parallel}_{d_{n},d_{n}} = \mathcal{I}^{\parallel,\psi}_{d_{n},d_{n}}$, and the measurement number required scales as $t^{-2}$ (Eq.\eqref{eq:QFI parallel static fields}). $\mathcal{I}^{\parallel}_{d_{n},d_{n}} = \mathcal{I}^{\parallel,\rho}_{d_{n},d_{n}}$ if the dynamical procedure is affected by noise, with the measurement that should be performed at $t=t^{\parallel, \rho}_{opt}$ to minimize the required measurements (Eq.\eqref{eq:max QFI parallel depolarized}). Finally, $\mathcal{I}^{\parallel}_{d_{n},d_{n}} = \mathcal{I}^{\parallel,\rho_{th}}_{d_{n},d_{n}}$ for estimation based on a thermal probe, where there is an optimal temperature for estimation $T=T^{\parallel}_{opt,\mathcal{F}_{\mathcal{E}}}$ (Eq.\eqref{eq:max FI thermal parallel}).

\subsection{Electron Electric Dipole Moment}

In the case of the electron the experimental results are based on measurement involving both the electron itself or molecules. In \cite{acme2018improved} the model correspond to an Hamiltonian in the form of Eq.\eqref{eq:EDM_Hamiltonian_red_parallel}, i.e. the parallel dipole moments configuration, as the physical properties of the system depends on the Larmor angular velocity 
\begin{equation}
    \label{eq:larmor electron parallel}
    \omega_{e} = \pm \mu \cdot B_{z} + d_{e} \cdot E_{z} \, .
\end{equation}
Following the same reasoning made for the neutron, in this condition the sloppiness of the model implies again that every experimental procedure aimed to characterize the dipolar properties of the model needs a precise prior knowledge of either the magnetic or electric dipole moment. Since the focus was on the estimation of the electron electric dipole moment the condition to fulfill is $\nicefrac{\Delta(d_{e})}{d_{e}} \ll 1$, leading to
\begin{equation}
    \label{eq:condition electron parallel}
    \mathsmaller{ \sqrt{ \left( \frac{ \Delta(\omega_{e}) + ( \mu \cdot \Delta(B_{z}) + B_{z} \cdot \Delta(\mu)) }{ \omega_{e} + \mu \cdot B_{z} } \right)^2 + \left( \frac{\Delta( E_{z})}{E_{z}} \right)^2  } \ll 1 }
\end{equation}
Again, using the definition of the (Q)FI we can derive the number of measurements required to achieve a precise value of the electric dipole moment of the electron
\begin{equation}
    \label{eq:measurement electron parallel}
    M^{\parallel}_{e} \gg \frac{1}{d_{e}^2 \cdot \mathcal{I}^{\parallel}_{d_{e},d_{e}} } \, ,
\end{equation}
where $\mathcal{I}^{\parallel}_{d_{e},d_{e}}$ can assume the values illustrated in Sec.\ref{sec:RaD} depending on the nature of the estimation procedure itself. As for the neutron for an ideal dynamical procedure the precision scales accordingly to Eq.\eqref{eq:QFI parallel static fields}, while in the presence of noise the measurement should be performed at $t=t^{\parallel, \rho}_{opt}$ to minimize the required measurements according to Eq.\eqref{eq:max QFI parallel depolarized}. Finally, for estimation based on a thermal probe, $\mathcal{I}^{\parallel}_{d_{n},d_{n}} = \mathcal{I}^{\parallel,\rho_{th}}_{d_{n},d_{n}}$ with the optimal temperature for estimation that is $T=T^{\parallel}_{opt,\mathcal{F}_{\mathcal{E}}}$ as expressed in Eq.\eqref{eq:max FI thermal parallel}. In a different configurational scenario \cite{vutha2011magnetic} it is possible to perform measurement of the permanent electric dipole moment of the electron according to the orthogonal dipole moments configuration. The physical properties of the system depends on the Larmor angular velocity 
\begin{equation}
    \label{eq:larmor electron orthogonal}
    \omega_{e} = \sqrt{ (\mu B_{x})^2 + (d_{e} E_{z})^{2} }  \, ,
\end{equation}
and together with the condition $\nicefrac{\Delta(d_{e})}{d_{e}} \ll 1$ this leads to
\begin{equation}
\label{eq:condition electron orthogonal}
    \mathsmaller{ \sqrt{ \left[ \frac{\sqrt{(2w \Delta w)^2 + (2 \mu B^2 \Delta \mu )^2 + (2 \mu ^2 B \Delta B)^2}}{2(w^2 - \mu ^2 B^2)} \right]^2 + \left( \frac{\Delta E}{E} \right)^2 } \ll 1 } \, .
\end{equation}
With the definition of the (Q)FI we can derive the number of measurements required to achieve a precise value of the electric dipole moment of the electron
\begin{equation}
    \label{eq:measurement electron orthogonal}
    M^{\perp}_{e} \gg \frac{1}{d_{e}^2 \cdot \mathcal{I}^{\perp}_{d_{e},d_{e}} } \, ,
\end{equation}
where again $\mathcal{I}^{\perp}_{d_{e},d_{e}}$ can assume the values illustrated in Sec.\ref{sec:RaD} depending on the nature of the estimation procedure itself. 

\subsection{Subatomic Multiparameter Estimation}

The results of Sec.\ref{sec:RaD} suggest that in the orthogonal dipole moments configuration the joint estimation of dipole moments is possible for the electron in appropriate experimental setup, such as \cite{vutha2011magnetic}, with the precision bounds that are defined by Eq.\eqref{eq:Bound Multiparameter R}. Following the same analysis done for the single parameter fundamental estimation, every reliable experiment should satisfy the condition $\sqrt{\left( \nicefrac{\Delta_{d}}{d^2} + \nicefrac{\Delta_{\mu}}{\mu^2} \right)} \ll 1$, where the relative errors of $\mu$ and $d$ follows equations in the form of Eq.\eqref{eq:condition electron orthogonal}. In the multiparameter estimation bounds formalism it is possible to relate the relative error condition to the QFI matrix \cite{previdi2026oh} assuming as a weight matrix $\boldsymbol{W}=diag(\frac{1}{d^2},\frac{1}{\mu^2})$, obtaining again a condition for the minimal number of measurement required to obtain solid experimental results. This condition reads
\begin{equation}
    \label{eq:Subatomic Multiparameter Estimation}
    M^{\perp}_{\mu-d} \gg \frac{ 1 + R }{\mathcal{I}_{d,d} \cdot \mathcal{I}_{\mu,\mu} - \mathcal{I}_{d,\mu} \cdot \mathcal{I}_{\mu,d}} \left( \frac{\mathcal{I}_{d,d}}{\mu^2} + \frac{\mathcal{I}_{\mu,\mu}}{d^2} \right) \, ,
\end{equation}
where the specific value of the QFI depends on the metrological procedure adopted (see Eqs.\eqref{eq:QFI orthogonal static fields diagonal elements},\eqref{eq:QFI orthogonal static fields noisy} and \eqref{eq:quantum_fisher_information_th_result_orthogonal}). The quantum multiparameter incompatibility term $ R= \sqrt{\nicefrac{\det(\boldsymbol{U})}{\det(\boldsymbol{\mathcal{I}})}} $, with $\boldsymbol{U}$ that follows Eqs. \eqref{eq:Uhlmann orthogonal},\eqref{eq:Uhlmann Matrix orthogonal field noisy} and \eqref{eq:Uhlmann Matrix orthogonal field thermal}, define the additional non commutativity term related to the incompatibility of the optimal measurements of the dipole moments. It assumes value $R=1$ for the dynamical case and $R=0$ for the static thermal probe, implying that in the static thermal condition the joint estimation of $\mu$ and $d$ is not influenced by any additional quantum noise related to the non commutativity of the symmetric logarithmic derivatives. Overall, we can conclude that for subatomic particles the simultaneous estimation of magnetic and electric dipole moments is possible, only if their Hamiltonian is in the form of Eq.\eqref{eq:EDM_Hamiltonian_red_orthogonal} and if the number of measurement adopted respect the condition of Eq.\eqref{eq:Subatomic Multiparameter Estimation}. 

\section{Conclusions} 
\label{sec:C}

In this paper we have investigated the ultimate bound on precision in the estimation of the electric and magnetic dipole moments of a quantum system through the analytical analysis of the quantum and classical Fisher information. Because of the symmetries of quantum mechanical systems, we have focused our analysis on two distinct configurations: the case in which the magnetic and the electric dipole moments are parallel and the case in which they are orthogonal. 

Under the assumption of the control of the system as a probe, we have identified the optimal initial pure state for quantum estimation in dynamic conditions and proposed a suitable measurement procedure for the characterization of the system. In the parallel dipole moments configuration we have found that in non noisy conditions the precision of an ideal electric or magnetic dipole measurement scales as $\propto t^2$, while in the noisy conditions there exist an ideal time in which the QFI of $\mu$ and $d$ is maximal. This configuration has an associated null Uhlmann matrix, and, in principle, a simultaneous estimation of both the dipole moment is possible without any additional quantum noise related to the non commutativity of the ideal measurement procedures of $\mu$ and $d$. Nonetheless, in this configuration the statistical model results sloppy, i.e. all the physical quantities associated to the system depends on a linear combination of $\mu$ and $d$ and so only a single parameter estimation of one of the two dipole components is possible with a priori knowledge of the other component. In the orthogonal dipole moments configuration the main results differs from the parallel case. In ideal conditions the precision of the dipole moment estimation includes a component $\propto t^2$ and an additional $\sin^2{(2 \omega_{\perp} t)}$ term; these define distinct metrological regimes depending on which contribution dominates. Considering a realistic noisy evolution the QFI is maximized for a specific time, which is different depending on which component dominates the QFI and on the dipole moment to be estimated. Differently from the parallel case the Uhlmann matrix is non null and then in a combined measurement of both dipole moments there is an intrinsic additional noise related to the non commutativity of the optimal measurement procedures of $\mu$ and $d$. Nonetheless, the determinant of the QFI matrix is time dependent and different from zero for all $t$ except that for $t^{*}=\nicefrac{n\pi}{2\omega_{\perp}}$ with $n\in \mathbb{N}$, meaning that the system depends on a linear combination of the dipole moments only for $t^{*}$. Overall, we can state that in this configuration a joint estimation of the electric and magnetic dipole moments of the quantum system is then possible.

Additionally, we have performed a theoretical analysis on a thermal probe, to study the effect of non ideal working condition on static metrological procedures. In this scenario we have found that there is an optimal temperature for the precise characterization of the dipole moments with an energy measurements, which corresponds to $T_{opt} \approx \nicefrac{\omega}{(1.1997 \cdot \kappa_{B})}$ for both the parallel and orthogonal dipole moments configuration \footnote{For the parallel case $\omega = \omega_{\parallel}$ and for the orthogonal case $\omega=\omega_{\perp}$}. In the parallel dipole moments configuration the QFI equals the FI of an energy measurement for both the dipole moments meaning that in principle a measurement of the population of the energy levels is optimal for the estimation of $\mu$ and $d$. Nevertheless, the model inherits the same statistical properties of the dynamical case and then even if the Uhlmann matrix results null the system is sloppy also in thermal configuration, implying that only a separate estimation of the dipole moments is possible. For the orthogonal dipole moments configuration the situation slightly differs. The ideal estimation procedures actually depends on the values of the dipoles themselves, and so also the position of the maxima of the diagonal elements of the QFI matrix. Indeed, in this configuration the FI of an energy measurement does not equals the QFI of the model, meaning that the measurement of the population of the energy levels is convenient but suboptimal for the estimation of the dipole moments. Additionally the FI matrix of an energy measurement has a null determinant, resulting in a sloppy procedure for the estimation of the dipoles.

Overall, this paper theoretically demonstrates when the precise characterization of the electric and magnetic dipole moment is possible both with dynamical and static methodologies and with pure or non pure probes. Our results define the optimal value of temperature or time for ideal static or dynamic characterization of quantum systems providing insights for the ultimate limits of precise characterization of magnetic and electric dipole moments. Indeed, despite the profound theoretical nature of this paper, it may find application in all experimental procedures aimed to estimate the electromagnetic properties of quantum systems, eventually contributing in the estimation of the neutron and electron dipole moments in both elementary particles (parallel dipole moments configuration) or different physical systems, such as $ThO$ or $YbF$ molecular compounds (orthogonal dipole moments configuration) or beyond, since the general theoretical model proposed in independent of the specific physical model to which it is applied.

\appendix

\section{Additional Material}
\label{app:AM}
We here present additional details of the local coordinate frame used for the formulation of the $\mathcal{H}_{r}$ and for the mathematical analysis of the quantum properties of Hamiltonians acting on two-dimensional Hilbert Spaces.

\subsection{Suitable Local Coordinate Frame}
\label{sec:AM1}
The basis of almost all the physical experiments relies in the definition of the most suitable reference frame for the analysis of the system under investigation \cite{swokowski1979calculus}. Any problem that involves a local electromagnetic field defined by $\vec{E} = (E_{x'}, E_{y'}, E_{z'})$ and $\vec{B}=(B_{x'}, B_{y'}, B_{z'})$ defined in a reference frame $\{x^{'} , y^{'} , z^{'}\}$. Without loss of generality we can consider a new reference frame which has as $z$ axis the versor
\begin{equation}
    \label{eq:new_z}
    \hat{z} = \hat{E} = \frac{\vec{E}}{\norm{\vec{E}}} = \frac{\vec{E}}{\sqrt{E_{x'}^2 + E_{y'}^2 + E_{z'}^2}},
\end{equation}
leading to an electric field defined as $\vec{E} = (0, 0, E_{z})$ with $E_{z} = \norm{\vec{E}}$. The magnetic field vector $\vec{B}$ have a component on the $z$ axis which corresponds to its projection on $\hat{E}$ defined as $\vec{B_{z}} = \vec{B} \cdot \hat{E}$.
The magnetic component $\vec{B_{z}}$ can have the same or opposite direction of $\vec{E}$ and thus $\vec{B_{z}} = \pm \norm{\vec{B_{z}}} \hat{z}$. The remaining direction of the new reference frame $\hat{x}$ and $\hat{y}$ have to be both orthogonal to $\hat{z}$ and among each other, i.e. they identify a $D=2$ surface whose normal is the direction $\hat{z}$. Without loss of generality it is convenient to define the quantity
\begin{equation}
    \label{eq:new_Bx}
    \norm{\vec{B_{x}}} = \sqrt{\norm{B}^2 - B_z^2}.
\end{equation}
which intrinsically defines the versor $\hat{x}$ as
\begin{equation}
    \label{eq:new_x}
    \hat{x} = \frac{\vec{B} - \vec{B_{z}}}{\norm{\vec{B} - \vec{B_{z}}}} = \frac{\vec{B_{x}}}{\norm{\vec{B_{x}}}}.
\end{equation}
The remaining coordinate versor $\hat{y}$, orthogonal to both $\hat{x}$ and $\hat{z}$, is obtained as $\hat{y} = \hat{z} \cross \hat{x}$.
Summarizing: the same vectors $\vec{E}$ and $\vec{B}$ can be represented in the two reference frames $\{x^{'} , y^{'} , z^{'}\}$ and $\{x , y , z\}$ as in Table  \ref{tab:noises}
\begin{table}[h!]
    \centering
    \begin{tabular}{|c|c|c|}
    \hline
       $ + $ & $\{x^{'} , y^{'} , z^{'}\}$ & $\{x , y , z\}$ \\ [0.5ex] 
       \hline
       
       $\vec{E}$ & $(E_{x'}, E_{y'}, E_{z'})$ & $(0,0,E_z)$ \\ [0.5ex] 
       \hline
       
       $\vec{B}$ & $(B_{x'}, B_{y'}, B_{z'})$ & $(B_x,0,\pm B_z)$ \\ [0.5ex] 
       \hline
       
    \end{tabular}
    \caption{Table that summarizes the passage from a general reference frame to a suitable reduced local coordinate frame.}
    \label{tab:noises}
\end{table}

\subsection{Two-Dimensional Hilbert Space Physics}
\label{sec:AM2}
In this subsection we report mathematical details useful for the calculation of the physical properties of any two level system, from the matrix exponentiation to the eigenproblem, two of the basic building blocks of the quantum mechanic nature of a system. 
\subsubsection{Matrix Exponentiation}
A general matrix $\Bar{M}$, in a two level Hilbert space, can be seen as a combination of Pauli and identity matrices, as
\begin{equation}
    \label{eq:matrix_decomposition}
    \Bar{M}= \mathbb{I}\xi_0 + \sum_{l=1}^{3} \xi_l \sigma_{l}
\end{equation}
with the identity term that results only in a shift in energy if $M$ is the Hamiltonian of the system, without changing the physics of the system, and thus can be neglected. Then, it is possible to consider an Hamiltonian in the form
\begin{equation}
    \label{eq:matrix_decomposition_red}
    M= \sum_{l=1}^{3} \xi_l \sigma_{l} = \Xi \left( \hat{n}\cdot \Vec{\sigma} \right),
\end{equation}
where $\Vec{\sigma}$ is the vector of Pauli matrices (see Eq.\eqref{eq:Pauli_Vector}), $\hat{n}$ is the versor of the coefficients $\xi_l$ and $\Xi = \sqrt{\sum_{l=1}^{3} \abs{\xi_l}^2}$. For the complete knowledge of the state of a quantum system at time $t$ it is necessary to pass from the Hamiltonian to the time evolution operator through the exponentiation of the matrix $M$. Following the expression Eq.\eqref{eq:matrix_decomposition_red}, the exponential of the matrix $M$ assume the form 
\begin{align}
    \label{eq:matrix_exp}
    &e^{i\Xi\left( \hat{n}\cdot \Vec{\sigma} \right)t} = \sum_{k=0}^{\infty} \frac{i^{k} \left[ \Xi \left( \hat{n}\cdot \Vec{\sigma} \right) \right]^k}{k!} \nonumber \\
    &=\mathbb{I}\cos{(\Xi)} + i\hat{n}\cdot \Vec{\sigma} \sin{(\Xi)} \, ,
\end{align}
and represent the general form of the time evolution operator of a bi-dimensional quantum state.

\subsubsection{The Eigenproblem}
Another fundamental aspect of quantum mechanics is the eigenvalue and eigenvector problem. For any $2\times2$ matrix in the form
\begin{equation}
    \label{eq:two_matrix}
    M=\left( \begin{array}{cc}
         \xi_3 & \xi_1-i\xi_2 \\
         \xi_1+i\xi_2  & -\xi_3
    \end{array} \right)
\end{equation}
the eigenproblem can be solved as
\begin{equation}
    \label{eq:eigenvalues_two_matrix}
    \begin{cases}
        \mathcal{E}_{0} = \frac{T}{2} - \left( \frac{T^2}{4} - D \right)^{\frac{1}{2}} \\
        \mathcal{E}_{1} = \frac{T}{2} + \left( \frac{T^2}{4} - D \right)^{\frac{1}{2}}
    \end{cases}
\end{equation}
where $T$ and $D$ correspond respectively to the trace and the determinant of the matrix $M$. Due to the traceless nature of the matrix presented in Eq.\eqref{eq:two_matrix} the eigenvalues can be rewritten as
\begin{equation}
    \label{eq:eigenvalues_two_matrix_reduced}
    \mathcal{E}_{\pm} = \pm \sqrt{-D} = \pm \abs{\Xi}
\end{equation}
The eigenvectors of such class of matrices are
\begin{align}
    \label{eq:eigenvectors_2}
    \ket{\phi_-}&= \phi_{0-} \ket{0} + \phi_{1-} \ket{1}  \nonumber \\
    &=\frac{1}{\sqrt{2\Xi(\Xi+\xi_3)}}\left( \begin{array}{cc}
         i\xi_2-\xi_1 \\
         \Xi + \xi_3
    \end{array} \right)
\end{align}
and
\begin{align}
    \label{eq:eigenvectors_1}
    \ket{\phi_+}&= \phi_{0+} \ket{0} + \phi_{1+} \ket{1}  \nonumber \\
    &=\frac{1}{\sqrt{2\Xi(\Xi+\xi_3)}}\left( \begin{array}{cc}
         \Xi + \xi_3 \\
         \xi_1+i\xi_2 
    \end{array} \right) \,.
\end{align}
With this states and energies it is possible to perform the time evolution of a general bi-dimensional quantum wavefunction without explicitly performing the matrix exponentiation of the Hamiltonian (see Eq.\eqref{eq:initial_state_time_evolution}).

\subsubsection{Optimal Dipoles Electro-magnetometry}

As the focus of the paper is the metrological characterization of a system with magnetic and electric dipole moments we devote the present section to the optimization of any dynamic metrological problem when both an electric and magnetic field are applied. As in Sec.\ref{sec:ToE}, without loss of generality, we consider the Hamiltonian that depends on 2 parameters $(\xi_1,\xi_3)$. The optimal state to perform a metrological analysis consist in the equal superposition of $\ket{\phi_+}$ and $\ket{\phi_-}$ \cite{pang2014quantum} as
\begin{align}
    \label{eq:ideal_state_phi}
    &\ket{\psi_{opt}}= \frac{1}{\sqrt{2}} \left( \ket{\phi_{-}} + e^{i\eta} \ket{\phi_{+}} \right) \\
    &= \frac{1}{\sqrt{2}} \left( (\phi_{0-}+e^{i\eta}\phi_{0+})\ket{0} + (\phi_{1-}+e^{i\eta}\phi_{1+})\ket{1} \right) \nonumber \\
    &= \frac{1}{\sqrt{2}} \left( \tau e^{i\epsilon}\ket{0} + \nu e^{i\delta}\ket{1} \right) \nonumber 
\end{align}
with $\phi_{0\pm} = \bra{0}\ket{\phi_{\pm}}$, $\phi_{1\pm} = \bra{1}\ket{\phi_{\pm}}$ and the modulus of $\tau$ and $\nu$ that are obtained as 
\begin{equation}
    \label{eq:moduli}
    \begin{cases}
       \abs{\tau}=\sqrt{\abs{\phi_{0-}+e^{i\eta}\phi_{0+}}^2} \\
       \abs{\nu}=\sqrt{\abs{\phi_{1-}+e^{i\eta}\phi_{1+}}^2}
    \end{cases}
\end{equation}
The normalization condition ensure that $(\abs{\tau}^2 + \abs{\nu}^2)/2 =1$. The optimal state for a general problem of electro-magnetometry setting $\eta=\pm\frac{\pi}{2}$, $\abs{\tau}=\abs{\nu}=1$, is expressed in the form
\begin{align}
    \label{eq:ideal_state_phases}
    &\ket{\psi_{opt}}= \frac{1}{\sqrt{2}} \left( \ket{0} + e^{i\theta_{\pm}} \ket{1} \right),
\end{align}
where $\theta_{\pm}$ refers to the phase that define the state depending on the choice of $\eta=\pm\frac{\pi}{2}$, and is obtained as
\begin{align}
    \label{eq:phase}
    \theta_{\pm}&= \arctan\left( \frac{\phi_{0-}}{\phi_{0+}} \right) - \arctan\left( \frac{\phi_{1-}}{\phi_{1+}} \right) \nonumber \\
    &=\delta - \epsilon = \pm \frac{\pi}{2}
\end{align}
Then, in any EM estimation problem, aimed to optimize the metrological properties of the system, the degrees of freedom of $\eta$ allows to consider a state in the form of Eq.\eqref{eq:ideal_state_phases}, with the phases $\theta_{\pm}$ intrinsically defined by the structure of the eigenvalues themselves.

\begin{acknowledgments}
This work has been done under the auspices of GNFM-INdAM.
\end{acknowledgments}

\appendix

\bibliography{EDMs.bib}

@article{paris2009quantum,
  title={Quantum estimation for quantum technology},
  author={Paris, Matteo G. A.},
  journal={International Journal of Quantum Information},
  volume={7},
  number={supp01},
  pages={125--137},
  year={2009},
  publisher={World Scientific}
}

@article{cavazzoni2024coin,
  title={Coin dimensionality as a resource in quantum metrology involving discrete-time quantum walks},
  author={Cavazzoni, Simone and Razzoli, Luca and Ragazzi, Giovanni and Bordone, Paolo and Paris, Matteo G. A.},
  journal={Physical Review A},
  volume={109},
  number={2},
  pages={022432},
  year={2024},
  publisher={APS}
}

@article{candeloro2024dimension,
  title={Dimension matters: precision and incompatibility in multi-parameter quantum estimation models},
  author={Candeloro, Alessandro and Pazhotan, Zahra and Paris, Matteo G. A.},
  journal={Quantum Science and Technology},
  volume={9},
  number={4},
  pages={045045},
  year={2024},
  publisher={IOP Publishing}
}

@article{albarelli2023fundamental,
  title={Fundamental limits of pulsed quantum light spectroscopy: Dipole moment estimation},
  author={Albarelli, Francesco and Bisketzi, Evangelia and Khan, Aiman and Datta, Animesh},
  journal={Physical Review A},
  volume={107},
  number={6},
  pages={062601},
  year={2023},
  publisher={APS}
}

@article{ragazzi2024generalized,
  title={Generalized phase estimation in noisy quantum gates},
  author={Ragazzi, Giovanni and Cavazzoni, Simone and Bordone, Paolo and Paris, Matteo G. A.},
  journal={Physical Review A},
  volume={110},
  number={5},
  pages={052425},
  year={2024},
  publisher={APS}
}

@article{forghieri2023quantum,
  title={Quantum estimation and remote charge sensing with a hole-spin qubit in silicon},
  author={Forghieri, Gaia and Secchi, Andrea and Bertoni, Andrea and Bordone, Paolo and Troiani, Filippo},
  journal={Physical Review Research},
  volume={5},
  number={4},
  pages={043159},
  year={2023},
  publisher={APS}
}

@article{secchi2025hole,
  title={Hole-spin qubits in germanium beyond the single-particle regime},
  author={Secchi, Andrea and Forghieri, Gaia and Bordone, Paolo and Loss, Daniel and Bosco, Stefano and Troiani, Filippo},
  journal={arXiv preprint arXiv:2505.02449},
  year={2025}
}

@article{fanucchi2025giant,
  title={Giant Rabi frequencies between qubit and excited hole states in silicon quantum dots},
  author={Fanucchi, Eleonora and Forghieri, Gaia and Secchi, Andrea and Bordone, Paolo and Troiani, Filippo},
  journal={Physical Review B},
  volume={111},
  number={20},
  pages={205409},
  year={2025},
  publisher={APS}
}

@article{razavian2019quantum,
  title={Quantum thermometry by single-qubit dephasing},
  author={Razavian, Sholeh and Benedetti, Claudia and Bina, Matteo and Akbari-Kourbolagh, Yahya and Paris, Matteo G. A.},
  journal={The European Physical Journal Plus},
  volume={134},
  number={6},
  pages={284},
  year={2019},
  publisher={Springer Berlin Heidelberg}
}

@article{daniotti2018qubit,
  title={Qubit systems subject to unbalanced random telegraph noise: quantum correlations, non-Markovianity and teleportation},
  author={Daniotti, Simone and Benedetti, Claudia and Paris, Matteo G. A.},
  journal={The European Physical Journal D},
  volume={72},
  number={12},
  pages={208},
  year={2018},
  publisher={Springer}
}

@article{baumgratz2016quantum,
  title={Quantum enhanced estimation of a multidimensional field},
  author={Baumgratz, Tillmann and Datta, Animesh},
  journal={Physical review letters},
  volume={116},
  number={3},
  pages={030801},
  year={2016},
  publisher={APS}
}

@article{liu2020quantum,
  title={Quantum Fisher information matrix and multiparameter estimation},
  author={Liu, Jing and Yuan, Haidong and Lu, Xiao-Ming and Wang, Xiaoguang},
  journal={Journal of Physics A: Mathematical and Theoretical},
  volume={53},
  number={2},
  pages={023001},
  year={2020},
  publisher={IOP Publishing}
}

@article{jing2015maximal,
  title={Maximal quantum Fisher information for general su (2) parametrization processes},
  author={Jing, Xiao-Xing and Liu, Jing and Xiong, Heng-Na and Wang, Xiaoguang},
  journal={Physical Review A},
  volume={92},
  number={1},
  pages={012312},
  year={2015},
  publisher={APS}
}

@article{brask2015improved,
  title={Improved quantum magnetometry beyond the standard quantum limit},
  author={Brask, Jonatan Bohr and Chaves, Rafael and Ko{\l}ody{\'n}ski, Janek},
  journal={Physical Review X},
  volume={5},
  number={3},
  pages={031010},
  year={2015},
  publisher={APS}
}

@article{abel2020measurement,
  title={Measurement of the permanent electric dipole moment of the neutron},
  author={Abel, Christopher and Afach, Samer and Ayres, Nicholas J and Baker, Colin A and Ban, Gilles and Bison, Georg and Bodek, Kazimierz and Bondar, Vira and Burghoff, Martin and Chanel, Estelle and others},
  journal={Physical Review Letters},
  volume={124},
  number={8},
  pages={081803},
  year={2020},
  publisher={APS}
}

@article{baker2006improved,
  title={Improved experimental limit on the electric dipole moment of the neutron},
  author={Baker, CA and Doyle, DD and Geltenbort, P and Green, K and Van der Grinten, MGD and Harris, PG and Iaydjiev, P and Ivanov, SN and May, DJR and Pendlebury, JM and others},
  journal={Physical review letters},
  volume={97},
  number={13},
  pages={131801},
  year={2006},
  publisher={APS}
}

@article{boixo2007generalized,
  title={Generalized limits for single-parameter quantum estimation},
  author={Boixo, Sergio and Flammia, Steven T and Caves, Carlton M and Geremia, John M},
  journal={Physical review letters},
  volume={98},
  number={9},
  pages={090401},
  year={2007},
  publisher={APS}
}

@article{pang2014quantum,
  title={Quantum metrology for a general Hamiltonian parameter},
  author={Pang, Shengshi and Brun, Todd A},
  journal={Physical Review A},
  volume={90},
  number={2},
  pages={022117},
  year={2014},
  publisher={APS}
}

@article{marvian2022operational,
  title={Operational interpretation of quantum fisher information in quantum thermodynamics},
  author={Marvian, Iman},
  journal={Physical Review Letters},
  volume={129},
  number={19},
  pages={190502},
  year={2022},
  publisher={APS}
}

@article{kammerlander2016coherence,
  title={Coherence and measurement in quantum thermodynamics},
  author={Kammerlander, Philipp and Anders, Janet},
  journal={Scientific reports},
  volume={6},
  number={1},
  pages={22174},
  year={2016},
  publisher={Nature Publishing Group UK London}
}

@article{gusarov2023optimized,
  title={Optimized emulation of quantum magnetometry via superconducting qubits},
  author={Gusarov, NN and Perelshtein, MR and Hakonen, PJ and Paraoanu, GS},
  journal={Physical Review A},
  volume={107},
  number={5},
  pages={052609},
  year={2023},
  publisher={APS}
}

@article{troiani2018universal,
  title={Universal quantum magnetometry with spin states at equilibrium},
  author={Troiani, Filippo and Paris, Matteo G. A.},
  journal={Physical review letters},
  volume={120},
  number={26},
  pages={260503},
  year={2018},
  publisher={APS}
}

@article{gutenkunst2007universally,
  title={Universally sloppy parameter sensitivities in systems biology models},
  author={Gutenkunst, Ryan N and Waterfall, Joshua J and Casey, Fergal P and Brown, Kevin S and Myers, Christopher R and Sethna, James P},
  journal={PLoS computational biology},
  volume={3},
  number={10},
  pages={e189},
  year={2007},
  publisher={Public Library of Science San Francisco, USA}
}

@article{daniels2008sloppiness,
  title={Sloppiness, robustness, and evolvability in systems biology},
  author={Daniels, Bryan C and Chen, Yan-Jiun and Sethna, James P and Gutenkunst, Ryan N and Myers, Christopher R},
  journal={Current opinion in biotechnology},
  volume={19},
  number={4},
  pages={389--395},
  year={2008},
  publisher={Elsevier}
}

@article{cavazzoni2024characterization,
  title={Characterization of partially accessible anisotropic spin chains in the presence of anti-symmetric exchange},
  author={Cavazzoni, Simone and Adani, Marco and Bordone, Paolo and Paris, Matteo G. A.},
  journal={New Journal of Physics},
  volume={26},
  number={5},
  pages={053024},
  year={2024},
  publisher={IOP Publishing}
}

@article{adani2024critical,
  title={Critical metrology of minimally accessible anisotropic spin chains},
  author={Adani, Marco and Cavazzoni, Simone and Teklu, Berihu and Bordone, Paolo and Paris, Matteo G. A.},
  journal={Scientific Reports},
  volume={14},
  number={1},
  pages={19933},
  year={2024},
  publisher={Nature Publishing Group UK London}
}

@book{swokowski1979calculus,
  title={Calculus with analytic geometry},
  author={Swokowski, Earl William},
  year={1979},
  publisher={Taylor \& Francis}
}

@article{vutha2011magnetic,
  title={Magnetic and electric dipole moments of the H 3 $\Delta$ 1 state in ThO},
  author={Vutha, Amar C and Spaun, Benjamin and Gurevich, Yulia Vsevolodovna and Hutzler, Nicholas Richard and Kirilov, Emil and Doyle, John M and Gabrielse, Gerald and DeMille, David},
  journal={Physical Review A—Atomic, Molecular, and Optical Physics},
  volume={84},
  number={3},
  pages={034502},
  year={2011},
  publisher={APS}
}

@article{commins2007electric,
  title={The electric dipole moment of the electron: An intuitive explanation for the evasion of Schiff’s theorem},
  author={Commins, Eugene D and Jackson, J David and DeMille, David P},
  journal={American Journal of Physics},
  volume={75},
  number={6},
  pages={532--536},
  year={2007},
  publisher={AIP Publishing}
}

@article{engel2025nuclear,
  title={Nuclear schiff moments and CP violation},
  author={Engel, Jonathan},
  journal={Annu. Rev. Nucl. Part. Sci},
  pages={129},
  volume={75},
  year={2025},
  publisher={Annual Reviews}
}

@article{acme2018improved, title={Improved limit on the electric dipole moment of the electron}, author={Andreev, V. and others}, volume={562}, ISSN={1476-4687}, url={http://dx.doi.org/10.1038/s41586-018-0599-8}, DOI={10.1038/s41586-018-0599-8}, number={7727}, journal={Nature}, publisher={Springer Science and Business Media LLC}, year={2018}, month=Oct, pages={355–360} }

@article{graner2017erratum,
  title={Erratum: Reduced limit on the permanent electric dipole moment of Hg 199 [phys. rev. lett. 116, 161601 (2016)]},
  author={Graner, B and Chen, Y and Lindahl, EG and Heckel, BR},
  journal={Physical Review Letters},
  volume={119},
  number={11},
  pages={119901},
  year={2017},
  publisher={APS}
}

@article{parker2015first,
  title={First measurement of the atomic electric dipole moment of Ra 225},
  author={Parker, RH and Dietrich, MR and Kalita, MR and Lemke, ND and Bailey, KG and Bishof, M and Greene, JP and Holt, RJ and Korsch, Wolfgang and Lu, Z-T and others},
  journal={Physical Review Letters},
  volume={114},
  number={23},
  pages={233002},
  year={2015},
  publisher={APS}
}

@article{bishof2016improved,
  title={Improved limit on the Ra 225 electric dipole moment},
  author={Bishof, Michael and Parker, Richard H and Bailey, Kevin G and Greene, John P and Holt, Roy J and Kalita, Mukut R and Korsch, Wolfgang and Lemke, Nathan D and Lu, Zheng-Tian and Mueller, Peter and others},
  journal={Physical Review C},
  volume={94},
  number={2},
  pages={025501},
  year={2016},
  publisher={APS}
}

@article{daffer2004depolarizing,
  title={Depolarizing channel as a completely positive map with memory},
  author={Daffer, Sonja and W{\'o}dkiewicz, Krzysztof and Cresser, James D and McIver, John K},
  journal={Physical Review A—Atomic, Molecular, and Optical Physics},
  volume={70},
  number={1},
  pages={010304},
  year={2004},
  publisher={APS}
}

@article{manzano2020short,
    author = {Manzano, Daniel},
    title = {A short introduction to the Lindblad master equation},
    journal = {AIP Advances},
    volume = {10},
    number = {2},
    pages = {025106},
    year = {2020},
    month = {02},
    issn = {2158-3226},
    doi = {10.1063/1.5115323},
    url = {https://doi.org/10.1063/1.5115323}
}

@article{cavazzoni2024optimizing,
  title={Optimizing topology for quantum probing with discrete-time quantum walks},
  author={Cavazzoni, Simone and Bordone, Paolo and Paris, Matteo G. A.},
  journal={AVS Quantum Science},
  volume={6},
  number={4},
  year={2024},
  publisher={AIP Publishing}
}

@article{hudson2002measurement,
  title={Measurement of the electron electric dipole moment using YbF molecules},
  author={Hudson, JJ and Sauer, BE and Tarbutt, MR and Hinds, EA},
  journal={Physical review letters},
  volume={89},
  number={2},
  pages={023003},
  year={2002},
  publisher={APS}
}

@article{lim2018laser,
  title={Laser cooled YbF molecules for measuring the electron’s electric dipole moment},
  author={Lim, J and Almond, JR and Trigatzis, MA and Devlin, JA and Fitch, NJ and Sauer, BE and Tarbutt, MR and Hinds, EA},
  journal={Physical review letters},
  volume={120},
  number={12},
  pages={123201},
  year={2018},
  publisher={APS}
}

@article{sharma2025mitigating,
  title={Mitigating sloppiness in joint estimation of successive squeezing parameters},
  author={Sharma, Priyanka and Olivares, Stefano and Mishra, Devendra Kumar and Paris, Matteo G. A.},
  journal={New Journal of Physics},
  volume={27},
  number={10},
  pages={104511},
  year={2025},
  publisher={IOP Publishing}
}

@article{pospelov2005electric,
  title={Electric dipole moments as probes of new physics},
  author={Pospelov, Maxim and Ritz, Adam},
  journal={Annals of physics},
  volume={318},
  number={1},
  pages={119--169},
  year={2005},
  publisher={Elsevier}
}

@article{razavian2020quantumness,
  title={On the quantumness of multiparameter estimation problems for qubit systems},
  author={Razavian, Sholeh and Paris, Matteo G. A. and Genoni, Marco G},
  journal={Entropy},
  volume={22},
  number={11},
  pages={1197},
  year={2020},
  publisher={MDPI}
}

@article{carollo2019quantumness,
  title={On quantumness in multi-parameter quantum estimation},
  author={Carollo, Angelo and Spagnolo, Bernardo and Dubkov, Alexander A and Valenti, Davide},
  journal={Journal of Statistical Mechanics: Theory and Experiment},
  volume={2019},
  number={9},
  pages={094010},
  year={2019},
  publisher={IOP Publishing and SISSA}
}

@article{pospelov2014ckm,
  title={CKM benchmarks for electron electric dipole moment experiments},
  author={Pospelov, Maxim and Ritz, Adam},
  journal={Physical Review D},
  volume={89},
  number={5},
  pages={056006},
  year={2014},
  publisher={APS}
}

@article{ng2022spectroscopy,
  title={Spectroscopy on the electron-electric-dipole-moment--sensitive states of ThF+},
  author={Ng, Kia Boon and Zhou, Yan and Cheng, Lan and Schlossberger, Noah and Park, Sun Yool and Roussy, Tanya S and Caldwell, Luke and Shagam, Yuval and Vigil, Antonio J and Cornell, Eric A and others},
  journal={Physical Review A},
  volume={105},
  number={2},
  pages={022823},
  year={2022},
  publisher={APS}
}

@article{caldwell2023systematic,
  title={Systematic and statistical uncertainty evaluation of the HfF+ electron electric dipole moment experiment},
  author={Caldwell, Luke and Roussy, Tanya S and Wright, Trevor and Cairncross, William B and Shagam, Yuval and Ng, Kia Boon and Schlossberger, Noah and Park, Sun Yool and Wang, Anzhou and Ye, Jun and others},
  journal={Physical Review A},
  volume={108},
  number={1},
  pages={012804},
  year={2023},
  publisher={APS}
}

@article{hubert2022electric,
  title={Electric dipole moments generated by nuclear Schiff moment interactions: A reassessment of the atoms Xe 129 and Hg 199 and the molecule TlF 205},
  author={Hubert, Micka{\"e}l and Fleig, Timo},
  journal={Physical Review A},
  volume={106},
  number={2},
  pages={022817},
  year={2022},
  publisher={APS}
}

@article{denis2019enhancement,
  title={Enhancement factor for the electric dipole moment of the electron in the BaOH and YbOH molecules},
  author={Denis, Malika and Haase, Pi AB and Timmermans, Rob GE and Eliav, Ephraim and Hutzler, Nicholas R and Borschevsky, Anastasia},
  journal={Physical Review A},
  volume={99},
  number={4},
  pages={042512},
  year={2019},
  publisher={APS}
}

@article{augenbraun2020laser,
  title={Laser-cooled polyatomic molecules for improved electron electric dipole moment searches},
  author={Augenbraun, Benjamin L and Lasner, Zack D and Frenett, Alexander and Sawaoka, Hiromitsu and Miller, Calder and Steimle, Timothy C and Doyle, John M},
  journal={New Journal of Physics},
  volume={22},
  number={2},
  pages={022003},
  year={2020},
  publisher={IOP Publishing}
}

@article{flambaum2020electric,
  title={Electric dipole moments of atoms and molecules produced by enhanced nuclear Schiff moments},
  author={Flambaum, VV and Dzuba, VA},
  journal={Physical Review A},
  volume={101},
  number={4},
  pages={042504},
  year={2020},
  publisher={APS}
}

@inproceedings{martin2020current,
  title={Current status of neutron electric dipole moment experiments},
  author={Martin, JW},
  booktitle={Journal of Physics: Conference Series},
  volume={1643},
  pages={012002},
  year={2020},
  organization={IOP Publishing}
}

@article{altarev2012next,
  title={A next generation measurement of the electric dipole moment of the neutron at the FRM II},
  author={Altarev, I and Beck, DH and Chesnevskaya, S and Chupp, T and Feldmeier, W and Fierlinger, P and Frei, A and Gutsmiedl, E and Kuchler, F and Link, P and others},
  journal={Nuovo Cimento-C},
  volume={35},
  number={4},
  pages={122},
  year={2012}
}

@article{piegsa2013new,
  title={New concept for a neutron electric dipole moment search using a pulsed beam},
  author={Piegsa, Florian M},
  journal={Physical Review C—Nuclear Physics},
  volume={88},
  number={4},
  pages={045502},
  year={2013},
  publisher={APS}
}

@article{pospelov2001neutron,
  title={Neutron electric dipole moment from electric and chromoelectric dipole moments of quarks},
  author={Pospelov, Maxim and Ritz, Adam},
  journal={Physical Review D},
  volume={63},
  number={7},
  pages={073015},
  year={2001},
  publisher={APS}
}

@article{pendlebury2015revised,
  title={Revised experimental upper limit on the electric dipole moment of the neutron},
  author={Pendlebury, JM and Afach, Samer and Ayres, Nicholas J and Baker, Charles A and Ban, Gilles and Bison, Georg and Bodek, Kazimierz and Burghoff, Martin and Geltenbort, Peter and Green, Katie and others},
  journal={Physical Review D},
  volume={92},
  number={9},
  pages={092003},
  year={2015},
  publisher={APS}
}

@book{minkin2012dipole,
  title={Dipole moments in organic chemistry},
  author={Minkin, Vladimir Isaakovich},
  year={2012},
  publisher={Springer Science \& Business Media}
}

@article{kotochigova2003ab,
  title={Ab initio calculation of the KRb dipole moments},
  author={Kotochigova, S and Julienne, Paul S and Tiesinga, Eite},
  journal={Physical Review A},
  volume={68},
  number={2},
  pages={022501},
  year={2003},
  publisher={APS}
}

@article{gubskaya2002total,
  title={The total molecular dipole moment for liquid water},
  author={Gubskaya, Anna V and Kusalik, Peter G},
  journal={The Journal of chemical physics},
  volume={117},
  number={11},
  pages={5290--5302},
  year={2002},
  publisher={American Institute of Physics}
}

@article{zhong2013fisher,
  title={Fisher information under decoherence in Bloch representation},
  author={Zhong, Wei and Sun, Zhe and Ma, Jian and Wang, Xiaoguang and Nori, Franco},
  journal={Physical Review A},
  volume={87},
  number={2},
  pages={022337},
  year={2013},
  publisher={APS}
}

@article{chapeau2015optimized,
  title={Optimized probing states for qubit phase estimation with general quantum noise},
  author={Chapeau-Blondeau, Fran{\c{c}}ois},
  journal={Physical Review A},
  volume={91},
  number={5},
  pages={052310},
  year={2015},
  publisher={APS}
}

@article{previdi2026oh,
  title={OH molecule as a quantum probe to jointly estimate electric and magnetic fields},
  author={Previdi, Luca and Albarelli, Francesco and Paris, Matteo G. A.},
  journal={Physical Review A},
  volume={113},
  number={6},
  pages={062612},
  year={2026},
  publisher={APS}
}

@article{pang2017optimal,
  title={Optimal adaptive control for quantum metrology with time-dependent Hamiltonians},
  author={Pang, Shengshi and Jordan, Andrew N},
  journal={Nature communications},
  volume={8},
  number={1},
  pages={14695},
  year={2017},
  publisher={Nature Publishing Group UK London}
}

@article{yang2022variational,
  title={Variational principle for optimal quantum controls in quantum metrology},
  author={Yang, Jing and Pang, Shengshi and Chen, Zekai and Jordan, Andrew N and Del Campo, Adolfo},
  journal={Physical Review Letters},
  volume={128},
  number={16},
  pages={160505},
  year={2022},
  publisher={APS}
}

@article{albarelli2017ultimate,
  title={Ultimate limits for quantum magnetometry via time-continuous measurements},
  author={Albarelli, Francesco and Rossi, Matteo AC and Paris, Matteo G. A. and Genoni, Marco G},
  journal={New Journal of Physics},
  volume={19},
  number={12},
  pages={123011},
  year={2017},
  publisher={IOP Publishing}
}

@article{rossi2020noisy,
  title={Noisy quantum metrology enhanced by continuous nondemolition measurement},
  author={Rossi, Matteo AC and Albarelli, Francesco and Tamascelli, Dario and Genoni, Marco G},
  journal={Physical Review Letters},
  volume={125},
  number={20},
  pages={200505},
  year={2020},
  publisher={APS}
}

@article{wildermuth2005microscopic,
  title={Microscopic magnetic-field imaging},
  author={Wildermuth, Stephan and Hofferberth, Sebastian and Lesanovsky, Igor and Haller, Elmar and Andersson, L Mauritz and Groth, S{\"o}nke and Bar-Joseph, Israel and Kr{\"u}ger, Peter and Schmiedmayer, J{\"o}rg},
  journal={Nature},
  volume={435},
  number={7041},
  pages={440--440},
  year={2005},
  publisher={Nature Publishing Group UK London}
}

@article{montenegro2022sequential,
  title={Sequential measurements for quantum-enhanced magnetometry in spin chain probes},
  author={Montenegro, Victor and Jones, Gareth Si{\^o}n and Bose, Sougato and Bayat, Abolfazl},
  journal={Physical Review Letters},
  volume={129},
  number={12},
  pages={120503},
  year={2022},
  publisher={APS}
}

@article{de2011single,
  title={Single-spin magnetometry with multipulse sensing sequences},
  author={de Lange, Ger and Rist{\`e}, D and Dobrovitski, VV and Hanson, R},
  journal={Physical review letters},
  volume={106},
  number={8},
  pages={080802},
  year={2011},
  publisher={APS}
}

@article{rodriguez2018probing,
  title={Probing one-dimensional systems via noise magnetometry with single spin qubits},
  author={Rodriguez-Nieva, Joaquin F and Agarwal, Kartiek and Giamarchi, Thierry and Halperin, Bertrand I and Lukin, Mikhail D and Demler, Eugene},
  journal={Physical Review B},
  volume={98},
  number={19},
  pages={195433},
  year={2018},
  publisher={APS}
}

@article{suzuki2015parameter,
  title={Parameter estimation of qubit states with unknown phase parameter},
  author={Suzuki, Jun},
  journal={International Journal of Quantum Information},
  volume={13},
  number={01},
  pages={1450044},
  year={2015},
  publisher={World Scientific}
}

@article{suzuki2016explicit,
  title={Explicit formula for the Holevo bound for two-parameter qubit-state estimation problem},
  author={Suzuki, Jun},
  journal={Journal of Mathematical Physics},
  volume={57},
  number={4},
  year={2016},
  publisher={AIP Publishing}
}

\end{document}